\documentclass[fontsize=11pt,parskip=half,DIV=10,toc=bib]{scrbook}
\usepackage[utf8]{inputenc}
\usepackage[T1]{fontenc}
\usepackage[ngerman]{babel}
\usepackage{amsmath}
\usepackage{amsfonts}
\usepackage{amssymb}
\usepackage{makeidx}
\usepackage{graphicx}
\usepackage{mathptmx}
 \usepackage[scaled=.90]{helvet}
 \usepackage{courier}
 \usepackage{caption}
\usepackage{subfigure}
 \usepackage[left=2cm,right=2cm,top=2cm,bottom=3cm]{geometry}
\usepackage{xspace}

\usepackage{graphicx}
\usepackage{txfonts}

\newcommand{\I}{\mathrm{i}}
\newcommand{\E}{\mathrm{e}}
\newcommand{\D}{\mathrm{d}}
\newcommand{\taudom}{\tau\delta\omega}

\begin{document}

\thispagestyle{empty}
\begin{center}

\Large\textbf{Fakult\"at f\"ur  Physik und Astronomie\\
Universit\"at Heidelberg}

\vspace{8cm}
\Large\textbf{Freier Induktionszerfall und Linienform\\
 im dreidimensionalen magnetischen Dipolfeld}
\vspace{6cm}

\normalsize
Bachelorarbeit in Physik\\
eingereicht von\\
\vspace{0.5cm}
\Large\textbf{Lukas R. Buschle}\\
\normalsize
\vspace{0.5cm}
geboren in Tuttlingen (Deutschland)\\
\vspace{0.5cm}
\Large\textbf{M\"arz 2015}
\normalsize

\newpage

\vspace{18cm}

Diese Bachelorarbeit wurde von Dr. Dr. Christian H. Ziener,\\
Deutsches Krebsforschungszentrum (DKFZ), Abteilung Radiologie E010 begleitet und\\
von Prof. Dr. Peter Bachert betreut.
\end{center}
\newpage

\Large\textbf{Abstract}\\
\normalsize
In this work, the time evolution of the free induction decay caused by the local dipole field of a spherical magnetic perturber is analyzed. The complicated treatment of the diffusion process is considered by the strong-collision approximation that allows a determination of the free induction decay in dependence of the underlying microscopic tissue parameters such as diffusion coefficient, sphere radius and susceptibility difference. The interplay between susceptibility- and diffusion-mediated effects yields several dephasing regimes of which, so far, only the classical regimes of motional narrowing and static dephasing for dominant and negligible diffusion, respectively, were extensively examined. Due to the asymmetric form of the three-dimensional dipole field for spherical objects, the free induction decay exhibits a complex component in contradiction to the cylindrical case, where the symmetric local two-dimensional dipole field only causes a purely real induction decay. Knowledge of the shape of the corresponding frequency distribution is necessary for the evaluation of more sophisticated pulse sequences and a detailed understanding of the off-resonance distribution allows improved quantification of transverse relaxation.

\vspace{3cm}
\Large\textbf{Abstrakt}\\
\normalsize
In dieser Arbeit wird die Zeitentwicklung des freien Induktionszerfalls analysiert, die durch ein lokales magnetisches Dipolfeld einer sphärischen, magnetischen Suszeptibilitätsinhomogenität induziert wird. Die komplexe Behandlung des Diffusionsprozesses wird durch die Strong-Collision-Näherung ersetzt, wodurch der freie Induktionszerfall in Abhängigkeit der zugrundeliegenden mikroskopischen Gewebeparametern, wie dem Diffusionskoeffizienten, dem Kugelradius und dem Suszeptibilitätsunterschied, bestimmt werden kann. Das Zusammenspiel von suszeptibilitäts- und diffusionsbezogenen Effekten verursacht mehrere Dephasierungsregime, von welchen bisher nur die klassischen Regime des Motional-Narrowing- und Static-Dephasing-Grenzfalls für dominante bzw. vernachlässigbare Diffusion ausführlich untersucht wurden. Aufgrund der asymmetrischen Form des dreidimensionalen Dipolfelds von sphärischen Objekten weist der freie Induktionszerfall eine komplexe Magnetisierung auf. Dies steht im Gegensatz zum zylindrischen Fall, bei dem das symmetrische lokale  zweidimensionale Dipolfeld einen rein reellen Zerfall bedingt. Für die Beurteilung von komplexen Pulssequenzen ist die Kenntnis über die Form der zugehörigen Frequenzverteilung notwendig. Zudem erlaubt ein detailliertes Verständnis der Off-Resonanz-Verteilung eine verbesserte Quantifizierung der transversalen Relaxation.

\tableofcontents
\newpage

\chapter{Einleitung}
\label{intro}
In der Magnetresonanztomographie ist die transversale Relaxation unter anderem von mikroskopischen Eigenschaften des untersuchten Gewebes abhängig. Dieser Effekt wird beispielsweise bei Signalveränderungen aufgrund von Eisenablagerungen bei neurodegenerativen Erkrankungen \cite{Sedlacik13} oder bei der Anwendung von Eisenoxid-Kontrastmitteln beobachtet und genutzt \cite{Kurz14}. In der funktionellen Magnetresonanzbildgebung werden die magnetischen Eigenschaften des Bluts genutzt, um Oxigenierungsgrade zu quantifizieren (BOLD-Effekt = Blood Oxygenation Level Dependent) \cite{Ogawa90}.
Das grundlegende physikalische Prinzip dieser Anwendungen ist der Suszeptibilitätsunterschied zwischen mikroskopischen magnetischen Objekten und dem umgebenden Gewebe, das die Spin-tragenden Teilchen beinhaltet. Insbesondere jedoch im Fall von kleinen magnetischen Objekten ist der Effekt der Diffusion von Spin-tragenden Teilchen auf die Signalentstehung nicht vernachlässigbar.\\
Suszeptibilitäts- und Diffusionseffekt haben auf die Signalentstehung einen gegensätzlichen Einfluss: Während die Linienverbreiterung des NMR-Signals auf Suszeptibilitätsunterschieden basiert, führt der Einfluss der Diffusion zu einer sich verengenden Frequenzverteilung \cite{Jackson99}.
Ein grundlegendes Verständnis des Zusammenspiels von suszeptibilitäts- und diffusionsabhängigen Prozessen hilft die Eigenschaften von mikroskopischen, magnetischen Partikeln innerhalb eines Gewebes zu quantifizieren.
Die durch den Suszeptibilitätsunterschied induzierte magnetische Feldinhomogenität kann durch eine Multipolentwicklung beschrieben werden, wobei der Dipolanteil in dieser Entwicklung dominiert \cite{Jackson99}. 

Im Grenzfall von vernachlässigbarer Diffusion, auch als Static-Dephasing-Grenzfall bekannt, wurde der freie Induktionszerfall  als Erstes von Brown für dreidimensionale Dipolfelder untersucht \cite{Brown61}. Später analysierten Yablonskiy und Haacke \cite{Yablonskiy94} den freien Induktionszerfall und die zugehörigen Relaxationszeiten um zylindrische und sphärische Objekte. Die Anwendung des Konzepts der Zustandsdichte, bekannt aus der Statistischen Physik, erlaubt es, die Frequenzverteilung der lokalen Larmor-Frequenzen um die magnetischen Objekte zu studieren. Der erste Schritt bei der Analyse solcher Frequenzverteilungen wurde im Zusammenhang mit Bildgebung an der Lunge gemacht. Dabei wurde die Frequenzverteilung mit der Histogramm-Methode durch Unterteilung des dephasierenden Volumens in Subvoxel bestimmt \cite{Case87}. Eine detaillierte Untersuchung der Dipolfeld-Signalentstehung von sphärischen magnetischen Feldinhomogenitäten wurde von Cheng \emph{et al.} durchgeführt, wobei ein analytischer Ausdruck für die Frequenzverteilung gefunden werden konnte \cite{Cheng01}. Weitere experimentellen Beweise für die Frequenzverteilung im Static-Dephasing-Regime können in \cite{Seppenwoolde05} gefunden werden. Objekte mit komplexerer Struktur werden in \cite{Durney89} betrachtet. Weitere Ergebnisse werden in \cite{Cutillo96} zusammengefasst. Die theoretischen Ergebnisse konnten experimentell bestätigt werden, insbesondere bei Lungengewebe, wo die grundlegende geometrische Struktur der dreidimensionale Dipolfelder durch luftgefüllte Alveolen (Lungenbläschen) erzeugt wird \cite{Ailion92,Christman96}. 

Neuerdings trat die Untersuchung von zylindrischen Objekten und deren Einfluss auf die Signalentstehung für die kardiale Bildgebung in den Fokus des Interesses, da das Myokard (Herzmuskel) hauptsächlich aus Kapillaren besteht, die aufgrund der paramagnetischen Eigenschaften von sauerstoffarmem Hämoglobin ein zweidimensionales Dipolfeld induzieren \cite{Bauer99a}. Durch Benutzung der bekannten Methoden für sphärische Feldinhomogenitäten war es analog möglich, die Frequenzverteilung um ein zylindrisches Objekt theoretisch \cite{Ziener05b} und experimentell \cite{Ziener07a,Sedlacik07} zu bestimmen.

Um diffusionsabhängige Signalzerfälle und deren Relaxationsprozesse zu verstehen, ist es essentiell, den Einfluss der Diffusion auf die Linienform der Frequenzverteilung zu analysieren. Im Allgemeinen kann die Kombination aus Suszeptibilitätseffekten und Diffusionseffekten auf die transversale Magnetisierung durch die Bloch-Torrey-Gleichung beschrieben werden \cite{Torrey56}. Ein erster Schritt zur Quantifizierung von diffusionsbezogenen Effekten auf die Signalform wurde von Bauer \emph{et al.} realisiert. Hierbei wurde die Strong-Collision-Näherung durch Ersetzen des Diffusionsoperators durch einen einfacheren stochastischen Prozess etabliert \cite{Bauer99a,Bauer99b}. Das Zusammenfügen von Strong-Collision-Näherung und dem Konzept der Zustandsdichte erlaubt es, den Effekt der Diffusion auf die Linienform und den freien Induktionszerfall zu bestimmen. Für zylindrische, magnetische Objekte weist die Frequenzverteilung ein typisches Muster mit zwei symmetrischen Peaks im Static-Dephasing-Regime auf. Mit zunehmender Diffusion verschmelzen diese Peaks zu einer Lorentz-Kurve \cite{Ziener07a}. Die symmetrischen Peaks verursachen Schwebungsfrequenzen, die vom harmonischen Oszillator bekannt sind \cite{Ziener12b}. Diese erlauben eine Klassifizierung von Diffusionsregimen in oszillierende oder zerfallende Zeitentwicklungen des freien Induktionszerfalls. Dies korrespondiert mit der Existenz von einem oder zwei Peaks in der Frequenzverteilung.

Aufgrund der komplexen Natur des dreidimensionalen Dipolfelds ist die Frequenzverteilung von sphärischen Objekten schwieriger zu erhalten. Jedoch wurden auch hier ähnliche Effekte wie Linienverschmälerung bei zunehmender Diffusion beobachtet (siehe Abb. 7 in \cite{Ziener07a}). Dies geht mit einem Übergang von einer asymmetrischen Frequenzverteilung im Static-Dephasing-Regime zu einer symmetrischen Frequenzverteilung im Motional-Narrowing-Regime einher. Ebenso wurden in Experimenten von Mulkern \emph{et al.} asymmetrische Kurvenformen der Frequenzverteilung für Lungengewebe gefunden \cite{Mulkern14}. 
Es stellt sich somit die Frage, ob für Diffusionsprozesse im dreidimensionalen Dipolfeld ebenfalls Analogien zum harmonischen Oszillator gefunden werden können. Des Weiteren ist das oszillierende oder zerfallende Verhalten des freien Induktionszerfalls für die quantitative Analyse und Kombination von suszeptibilitäts- und diffusionsgewichteter Bildgebung interessant.

In der vorliegenden Arbeit wird der Effekt der Diffusion auf den Dephasierungsprozess in einem dreidimensionalen Dipolfeld eines sphärischen magnetischen Objektes untersucht und der freie Induktionszerfall eines solchen Modells berechnet. Die Ergebnisse erlauben es, eine klare Unterteilung in Diffusionsregime vorzunehmen. Zudem erlaubt die charakteristische Form der Frequenzverteilung und des freien Induktionszerfalls mikroskopische Strukturen zu quantifizieren, die aus sphärischen magnetischen Komponenten aufgebaut sind.

Im  Abschnitt "`Methoden"'  werden vorangegangene Ergebnisse dargestellt und der Ansatz zur Berechnung des freien Induktionszerfalls formuliert. Im Abschnitt "`Ergebnisse"'  wird das Verhalten des freien Induktionszerfalls innerhalb verschiedener Regime beschrieben und der Ansatz zur Modellierung des Gewebes wird durch den Vergleich von theoretischen Ergebnissen mit experimentellen Daten gerechtfertigt. Im abschließenden Abschnitt "`Diskussion"'  findet eine Diskussion der Ergebnisse statt und es wird ein Ausblick auf die Analysemöglichkeiten von mikroskopischen Gewebeparametern gegeben. Im Anhang \ref{AnhangA} befinden sich Hintergrundinformationen zum Static-Dephasing-Grenzfall. Anhang \ref{AnhangB} fasst die Anwendung der Strong-Collision-Näherung aus vorangegangen Arbeiten zusammen. In Anhang \ref{AnhangC} findet sich die mathematische Herleitung des freien Induktionszerfalls wieder.

\chapter{Methoden}
\section{Diffusion und Dephasierung in einem dreidimensionalen Dipolfeld}
\begin{figure}
\begin{center}
\includegraphics[width=9 cm]{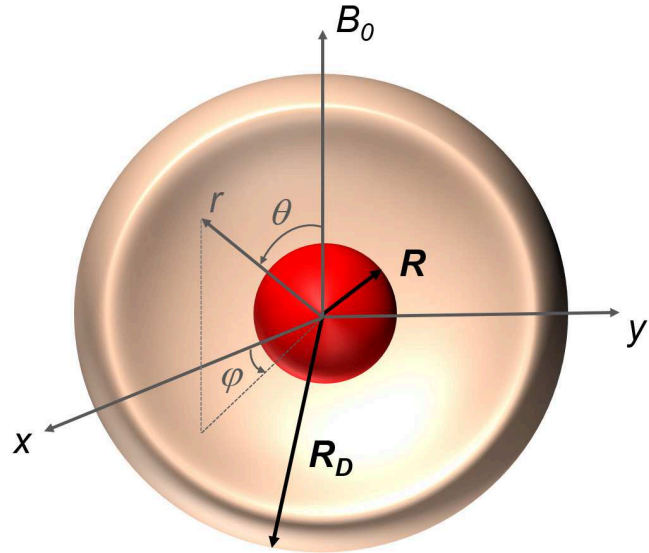}
\caption{
Magnetisierte Kugel mit Radius $R$, die von einem sphärischen Dephasierungsvolumen mit Radius $R_{\text{D}}$ umgeben ist, in dem die Diffusion und die Dephasierung der Spin-tragenden Teilchen auftritt. Das dreidimensionale Dipolfeld um die innere, magnetisierte Kugel wird durch den Abstand $r$ vom Ursprung und dem Winkel $\theta$ zum externen magnetischen Feld $B_0$ beschrieben.
}
\label{fig:Krogh}
\end{center}
\end{figure}

Wir betrachten eine Kugel mit Radius $R$, die von einem sphärischen Dephasierungsvolumen mit Radius $R_{\text{D}}$ umgeben ist (siehe Abb. \ref{fig:Krogh}). Die innere Kugel sei homogen magnetisiert und produziere ein dreidimensionales magnetisches Dipolfeld in der Form
\begin{equation}
\label{eq:frequenz}
\omega(\mathbf{r}) = \omega(r,\theta,\varphi) = \delta \omega R^3 \, \frac{3\cos^2(\theta)-1}{r^3} \,.
\end{equation}

Dabei bezeichnet  $\delta\omega=|\omega(r=R,\theta=\pi/2)|=\gamma B_{\text{äq}}$ die charakteristische äquatoriale Frequenzverschiebung und der Winkel $\theta$ wird zwischen dem externen magnetischen Feld $B_0$ und dem Positionsvektor $\mathbf{r}=(r,\theta,\varphi)$ gemessen.  Wie in Abb. \ref{fig:Krogh} dargestellt ist, wird der Positionsvektor $\mathbf{r}$ in sphärischen Koordinaten angegeben. Mit $\gamma$ wird das gyromagnetische Verhältnis des Protons bezeichnet. Das äquatoriale magnetische Feld $B_{\text{äq}}$ kann zu $B_{\text{äq}}=\mu_0\Delta M/3$ bestimmt werden, wobei $\mu_0 = 4\pi \times 10^{-7} \text{ kg m } \text{A}^{-2}\, \text{s}^{-2}$ die Vakuumpermeabilität und $\Delta M$ den Unterschied in der Magnetisierung zwischen der homogen magnetisierten inneren Kugel und dem umgebenden Gewebe bezeichnet. Es wird angenommen, dass das betrachtete sphärische magnetische Objekt nicht zur Signalbildung beiträgt. Spin-tragende Teilchen diffundieren in der sphärischen Schale $R\leq r\leq R_{\text{D}}$ um die lokale Feldinhomogenität. Dieser Diffusionsprozess wird durch den Diffusionskoeffizienten $D$ beschrieben. Das Volumenverhältnis $\eta$ ist definiert als: 
\begin{equation}
\label{eq:eta}
\eta = \frac{R^3}{R_{\text{D}}^3}\,.
\end{equation}

Die Annahme von reflektierenden Randbedingungen an der Oberfläche der äußeren Kugelschale erlaubt es, die periodische Struktur des beschriebenen Aufbaus innerhalb des Gewebes zu kompensieren \cite{Ziener08}. In diesem Fall ist der Diffusionsprozess durch die Korrelationszeit in der Mean-Relaxation-Time-Näherung charakterisiert \cite{Ziener06}.
\begin{equation}
\label{eq:tau}
\tau = \frac{R^2}{2\,D\,[1-\eta]}\,\left[1-\eta^{\frac{1}{3}}+\frac{4[1-\eta]^2+9 \left[2\eta - \eta^{\frac{5}{3}} -\eta^{\frac{1}{3}} \right]}{36 \left[\eta^{\frac{5}{3}}-1\right]}\right]
\end{equation}
Details zur Definition der Korrelationszeit $\tau$ finden sich in Anhang \ref{AnhangB} wieder.
Anschaulich beschreibt das Inverse der Korrelationszeit $1/\tau$ eine dynamische Frequenz, die mit der statischen Frequenz $\delta\omega$ verglichen werden kann. Für $1/\tau \gg \delta\omega$ überwiegen die Diffusionseffekte (Motional-Narrowing), für $1/\tau \ll \delta\omega$ ist der freie Induktionszerfall hauptsächlich durch den Suszeptibilitätsunterschied bestimmt (Static-Dephasing). Das Zusammenspiel dieser beiden Effekte wird in der vorliegenden Arbeit detailliert untersucht.

\section{Freier Induktionszerfall und Frequenzverteilung}

Im Allgemeinen wird die Zeitentwicklung der komplexwertigen lokalen  Magnetisierung $m(\mathbf{r},t)=m_x(\mathbf{r},t)+\I m_y(\mathbf{r},t)$ durch die Bloch-Torrey-Gleichung bestimmt \cite{Torrey56}:
\begin{align}
\label{eq:BT}
\frac{\partial}{\partial t} m(\mathbf{r},t) \, = \, \left[ D \Delta + \I \, \omega (\mathbf{r}) \right] \, m(\mathbf{r},t) \,,
\end{align}
wobei $\Delta$ den Laplace-Operator bezeichnet. Das gemessene Signal $M(t)$ setzt sich als Superposition der lokalen Magnetisierungen des Dephasierungsvolumens zusammen und lässt sich deshalb wie folgt schreiben:
\begin{align}
\label{eq:superposition}
M(t) = \frac{1}{V}\int_V\D^3{\mathbf r}\ m({\mathbf r},t)\,.
\end{align}
In der folgenden Analyse ist die intrinsische $T_2$-Relaxation zunächst vernachlässigt.
Da der gemessene freie Induktionszerfall $M(t)$ die transversalen Komponenten $M_x(t)$ und $M_y(t)$ enthält, die zu der komplexen Größe $M(t)=M_x(t) + \I M_y(t)$ addiert werden, ist es vorteilhaft, die reelle Frequenzverteilung $p(\omega)$ zu betrachten, die mit dem freien Induktionszerfall durch eine Fourier-Transformation verknüpft ist:
\begin{align} 
\label{FT-allg} 
M(t)  = \int\limits_{-\infty}^{+\infty} \text{d} \omega \,\, p(\omega) \, \text{e}^{{\text{i} \, \omega \, t}} \,.
\end{align}

\section{Static-Dephasing-Grenzfall}

Im Spezialfall für verschwindende Diffusion ($D=0$) wird die Zeitentwicklung der lokalen Magnetisierung $m_0(\mathbf{r},t)$ durch die Bloch-Gleichung $\partial_t m_0(\mathbf{r},t) = \I \omega (\mathbf{r}) m_0(\mathbf{r},t)$ beschrieben. Das zugehörige Diffusionsregime ist als Static-Dephasing-Regime bekannt, gekennzeichnet durch den Index $0$. In diesem Fall ist die Gesamtmagnetisierung des Dephasierungsvolumens $V=\frac{4}{3}\pi[R_{\text{D}}^3-R^3]$ gegeben durch \cite{Ziener07a}:
\begin{align}
\label{SignalSD} 
M_0(t) & = \frac{1}{V} \int_V \text{d}^3 \mathbf{r} \text{e}^{{\text{i} \, \omega(\mathbf{r}) \, t}} \\
\label{eq:M0Numerik}
&=\frac{1}{2} \frac{\eta}{1-\eta}\int\limits_{-1}^{+1} \! \D x \int\limits_{+1}^{\eta^{-1}} \! \D y  \E^{\I \delta\omega \frac{3x^2-1}{y}t}\\
\label{M01}
& = \frac{ h(\eta \, \delta\omega \, t) - \eta \, h(\delta\omega \, t)}{1-\eta}
\end{align} 
mit der kugelspezifischen $h$-Funktion, die in Gl. \eqref{app:kleinh} des Anhangs \ref{AnhangA} gegeben ist. Um den freien Induktionszerfall im Static-Dephasing-Grenzfall $M_0(t)$ numerisch zu berechnen, ist es sinnvoll Gl. \eqref{eq:M0Numerik} zu benutzen. Analog zu Gl. \eqref{FT-allg} kann das Static-Dephasing-Signal ebenfalls in Abhängigkeit der Frequenzverteilung $p_0(\omega)$ angegeben werden \cite{Cheng01,Ziener07a}:
\begin{equation} 
\label{FT} 
M_0(t)  = \int\limits_{-\delta\omega}^{+2\delta\omega} \text{d} \omega \,\, p_0(\omega) \, \text{e}^{{\text{i} \, \omega \, t}} \,,
\end{equation}
wobei die Frequenzverteilung des Static-Dephasing-Regime $p_0(\omega)$ in Gl. \eqref{eq:p0} in Anhang \ref{AnhangA} angegeben ist. Sie nimmt nur für $-\delta\omega \leq \omega \leq + 2\delta\omega$ von Null verschiedene Werte an.

\section{Erweiterung auf alle Diffusionsregime}
\begin{figure}
\begin{center}
\includegraphics[width=9 cm]{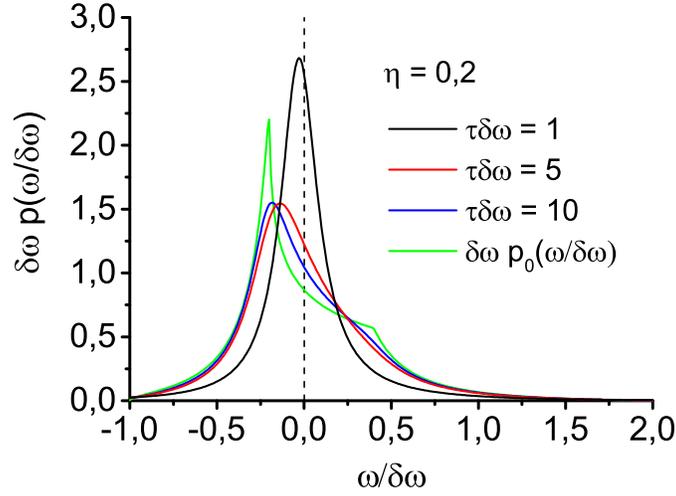}
\caption{\label{Fig:Fig2}
Frequenzverteilung $p(\omega)$ in Abhängigkeit der lokalen Resonanzfrequenz erhalten durch Gl. \eqref{eq:Fouriertransform} für verschiedene Werte des Parameters $\tau\delta\omega$  und einem Volumenverhältnis von $\eta=0,2$. Die Größen $\delta\omega p(\omega)$ auf der Ordinate und $\omega/\delta\omega$ auf der Abszisse sind dimensionslos. Im Static-Dephasing-Regime für den Grenzfall $\tau\delta\omega\to\infty$ (grüne Kurve) stimmt die Frequenzverteilung von Gl. \eqref{eq:p0} mit der Frequenzverteilung aus Gl. \eqref{eq:Fouriertransform} überein. }
\end{center}
\end{figure}

Auch wenn numerische Lösungen der Bloch-Torrey-Gleichung vorhanden sind \cite{Ziener09}, ist es mühsam analytische Lösungen zu finden, da die Gleichung einen nicht-hermiteschen Charakter aufweist. Eine angemessene Näherung um dieses Problem zu umgehen, ist das Einführen der Strong-Collision-Näherung. Es wurde gezeigt, dass diese Näherung über den gesamten Dynamikbereich, das heißt für Regime mit dominanten diffusionsabhängigen Beiträgen zur Signalform und für Regime mit vernachlässigbarer Diffusion gerechtfertigt ist  \cite{Bauer99a,Bauer99b,Bauer02}. Eine detaillierte Darstellung der Strong-Collision-Näherung, wie sie in vorangegangen Arbeiten eingeführt wurde, ist in Anhang \ref{AnhangB} gegeben. Ebenfalls konnte in einer vorangegangen Arbeit bereits eine explizite Frequenzverteilung für das dreidimensionale Dipolfeld angegeben werden, die für alle Diffusionsregime gültig ist \cite{Ziener07}:
\begin{align}
\label{eq:RueckFT}
p(\omega) & = \frac{1}{2\pi} \int\limits_{-\infty}^{+\infty} \text{d} t \,\, M(t) \, \text{e}^{{-\text{i} \, \omega \, t}} \\
\label{eq:Fouriertransform}
& = \frac{\tau}{\pi} \text{Re} \left(\! \frac{H \left( \frac{1+\I\tau\omega}{\eta\taudom}\right)-\eta H \left( \frac{1+\I\tau\omega}{\taudom} \right) }
{\left[ 1 - \eta \right] \left[1+\I\tau\omega \right]- H \left( \frac{1+\I\tau\omega}{\eta\taudom}\right)+\eta H \left( \frac{1+\I\tau\omega}{\taudom} \right)}\! \right)\\
\label{eq:FourierNenner}
& = \frac{\tau}{\pi} [1-\eta] \text{Re} \left(\frac{1+\I \tau\omega}{N(1+\I \tau\omega)} \right)-\frac{\tau}{\pi}
\end{align}
mit dem Nenner
\begin{equation}
\label{eq:Nenner}
N(s) = s[1-\eta] - H \left( \frac{s}{\eta \taudom} \right) + \eta H \left( \frac{s}{\taudom} \right)
\end{equation}
und mit der Funktion
\begin{equation}
\label{eq:H}
H(y) = \frac{1}{3}+\frac{2}{3} \left[ 1 - \frac{2 \I}{y}\right] \sqrt{\frac{1-\I y}{3}} \mathrm{arccoth} \left(\!\! \sqrt{\frac{1-\I y}{3}} \right)\,.
\end{equation}
Somit lässt sich die Frequenzverteilung $p(\omega)$  mittels Gl. \eqref{eq:Fouriertransform} bzw. Gl. \eqref{eq:FourierNenner} bestimmen. Mehrere Frequenzverteilungen sind in Abb. \ref{Fig:Fig2} für unterschiedliche Werte des Parameters $\taudom$ dargestellt. Für große Werte von $\taudom$ liegt das Static-Dephasing-Regime vor, in dem die Frequenzverteilung $p_0(\omega)$ in Gl. \eqref{eq:p0} des Anhangs \ref{AnhangA} gegeben ist (siehe auch grüne Linie in  Abb. \ref{Fig:Fig2}). In diesem Fall weist die Frequenzverteilung interessanterweise zwei auffällige Peaks bei den Frequenzen $\omega=-\eta\delta\omega$ und  $\omega=+2\eta\delta\omega$ auf, wie detailliert im Anhang \ref{AnhangA} in Gl. \eqref{peak1} und Gl. \eqref{peak2} dargestellt ist. 

Im Unterschied zum Static-Dephasing-Regime, in dem der Bereich der möglichen Resonanzfrequenzen auf das Intervall $-\delta\omega \leq \omega \leq + 2\delta\omega$ beschränkt ist, sind unter Berücksichtigung der Diffusion alle Resonanzfrequenzen möglich: $-\infty < \omega < +\infty$. Aufgrund dieser Eigenschaft ist die numerische Berechnung des Signals mittels Gl. \eqref{FT-allg} aufwändig. Deshalb ist es wünschenswert einen expliziten Ausdruck für das Signal $M(t)$ zu finden. 
 
Ein erster Schritt in dieser Analyse ist es, die Laplace-Transformation $\hat{M}(s)$ des freien Induktionszerfalls $M(t)$ zu betrachten, die mit dem Static-Dephasing-Regime über die Korrelationszeit (gegeben in Gl. \eqref{eq:tau}) verknüpft ist \cite{Ziener05a}, wie im Anhang \ref{AnhangB} ausgeführt wird:
\begin{align}
\label{eq:Laplacetransform}
\hat{M}(s) =& \int_0^\infty \D t \, \E^{-st}M(t)\,\\
\label{eq:LT}
=&\frac{\hat{M}_0(s+\tau^{-1})}{1-\tau^{-1}\hat{M}_0(s+\tau^{-1})}\,,
\end{align}
wobei die Laplace-Transformierte $\hat{M}_0(s) =\int_0^{\infty} \D t \, \E^{-st}M_0(t)$ der Magnetisierung $M_0(t)$ im Static-Dephasing-Regime aus Gl. \eqref{M01} folgt:
\begin{align}
\hat{M}_0(s) &= \frac{1}{1-\eta} \frac{1}{s} \left[ H \left( \frac{s}{\eta\delta\omega} \right) - \eta H \left( \frac{s}{\delta\omega} \right) \right]\\
&=1-\frac{N(s)}{s[1-\eta]}\,,
\end{align}
wobei die Funktion $H$ in Gl. \eqref{eq:H} und die Funktion $N$ in Gl. \eqref{eq:Nenner} gegeben ist. Gemäß ihrer Definitionen in Gl. \eqref{eq:RueckFT}  und Gl. \eqref{eq:Laplacetransform} sind Fourier-Transformation und Laplace-Transformation mittels $p(\omega) = \frac{1}{\pi} \text{Re}(\hat M(\I \omega))$ miteinander verknüpft.

Die einfachste anzunehmende Form des freien Induktionszerfalls ist ein monoexponentieller Zerfall \cite{Yung03}:
\begin{equation}
\label{eq:R2}
M(t)=\E^{-R_{2}^{\prime}t} \, .
\end{equation}
Die Relaxationsrate $R_{2}^{\prime}$ kann durch Benutzung der Mean-Relaxation-Time-Näherung bestimmt werden \cite{Nadler88}:
\begin{align}
\frac{1}{R_2^{\prime}} = \int\limits_0^{\infty} \frac{M(t)}{M(0)} \D t = \hat M(0)\,.
\end{align}
Dies ist eng mit der Laplace-Transformation der Magnetisierung aus Gl. \eqref{eq:LT} verbunden und kann deshalb folgendermaßen geschrieben werden:
\begin{align}
\label{R2strich-MRT}
R_2^{\prime} &= \frac{1}{\tau} \left[ \frac{1-\eta}{H\left(\frac{1}{\eta\taudom}\right)-\eta H\left(\frac{1}{\taudom}\right)} -1 \right] \\
& = \frac{1}{\tau}\frac{N(1)}{1-\eta-N(1)} \,.
\end{align}

Wie bereits beschrieben, ist insbesondere für große Werte des Parameters $\taudom$ der freie Induktionszerfall weder rein monoexponentiell noch rein reell.
Deshalb muss im Allgemeinen die Zeitentwicklung $M(t)$ durch Bestimmung der inversen Laplace-Transformation von $\hat{M}(s)$ berechnet werden.

Durch Kombination von allgemeinen Zeitskalierungs- und Dämpfungseigenschaften der Laplace-Transformation kann die Magnetisierung $M(t)$ umgeschrieben werden:
\begin{equation}
\label{eq:mf}
M(t) = \E^{-\frac{t}{\tau}} G \left(\frac{t}{\tau} \right)\,.
\end{equation}
Die Laplace-Transformierten $\hat{G}(s) = \int_0^{\infty} \mathrm{d}t \, \mathrm{e}^{-st} G(t)$ von $G(t)$ ist damit durch
\begin{align}
\hat{G}(s) & = \frac{1}{\tau} \hat{M} \left( \frac{s-1}{\tau} \right)\\
&= \frac{\tau^{-1}\hat M_0\left(\tau^{-1}s\right)}{1-\tau^{-1}\hat M_0\left(\tau^{-1}s\right)}\\
\label{eq:Gdach}
&= s\frac{1-\eta}{N(s)} - 1 
\end{align}
bestimmt, wobei der Nenner $N$ in Gl. \eqref{eq:Nenner} gegeben ist.

\begin{figure}
\begin{center}
\includegraphics[width=11cm ]{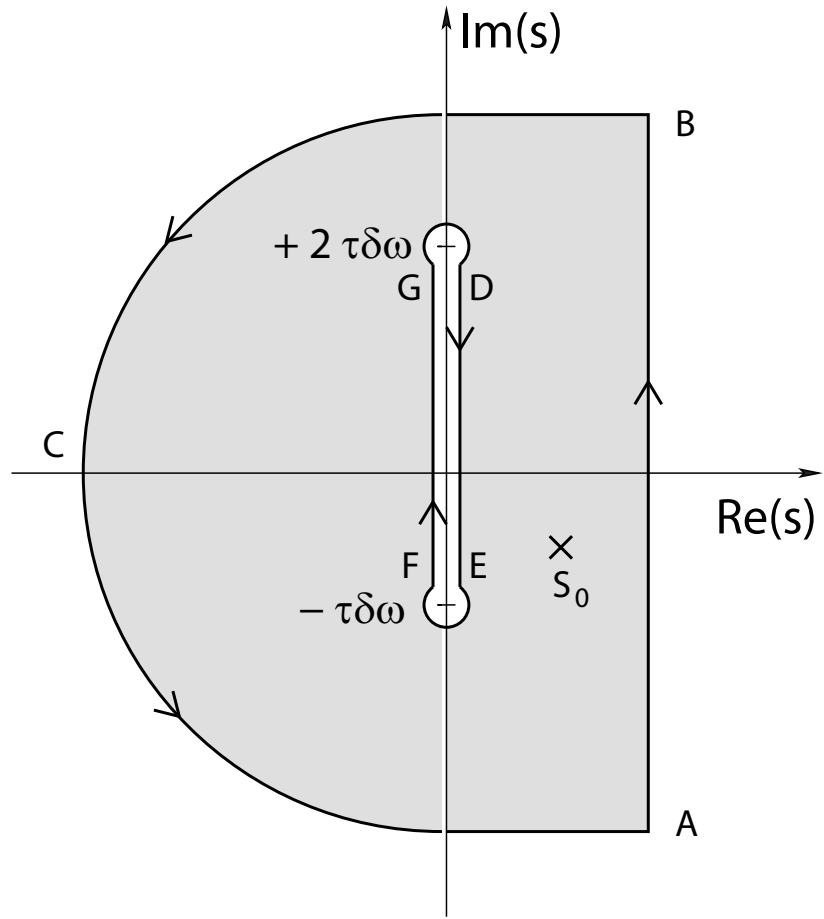}
\caption{ Integrationsweg in der komplexen $s$-Ebene. Die Verzweigungslinie $DEFGD$ der Funktion $\hat G(s)$ aus Gl. \eqref{eq:Gdach} ist auf der imaginären Achse im Bereich von $-\tau\delta\omega \leq \text{Im}(s) \leq +2 \tau \delta\omega $  lokalisiert. Die Singularität $s_0$ von $\hat G(s)$ liegt im rechten unteren Quadranten.}
\label{fig:Bromwich}
\end{center}
\end{figure}

Die inverse Laplace-Transformation der Funktion $\hat{G}(s)$ kann durch Anwendung der Mellin-Inversionsformel in der komplexen Ebene berechnet werden, wie in Abb. \ref{fig:Bromwich} dargestellt:
\begin{equation}
\label{eq:Bromwich}
G(t) = \frac{1}{2\pi\I} \int_A^B \hat{G}(s) \E^{st} \D s\,.
\end{equation}
Die Berechnung der inversen Laplace-Transformation wird detailliert in Anhang \ref{AnhangC} beschrieben.

\chapter{Ergebnisse}
\section{Klassifikation von Diffusionsregimen}
\begin{figure}
\begin{center}
\includegraphics[width=\textwidth]{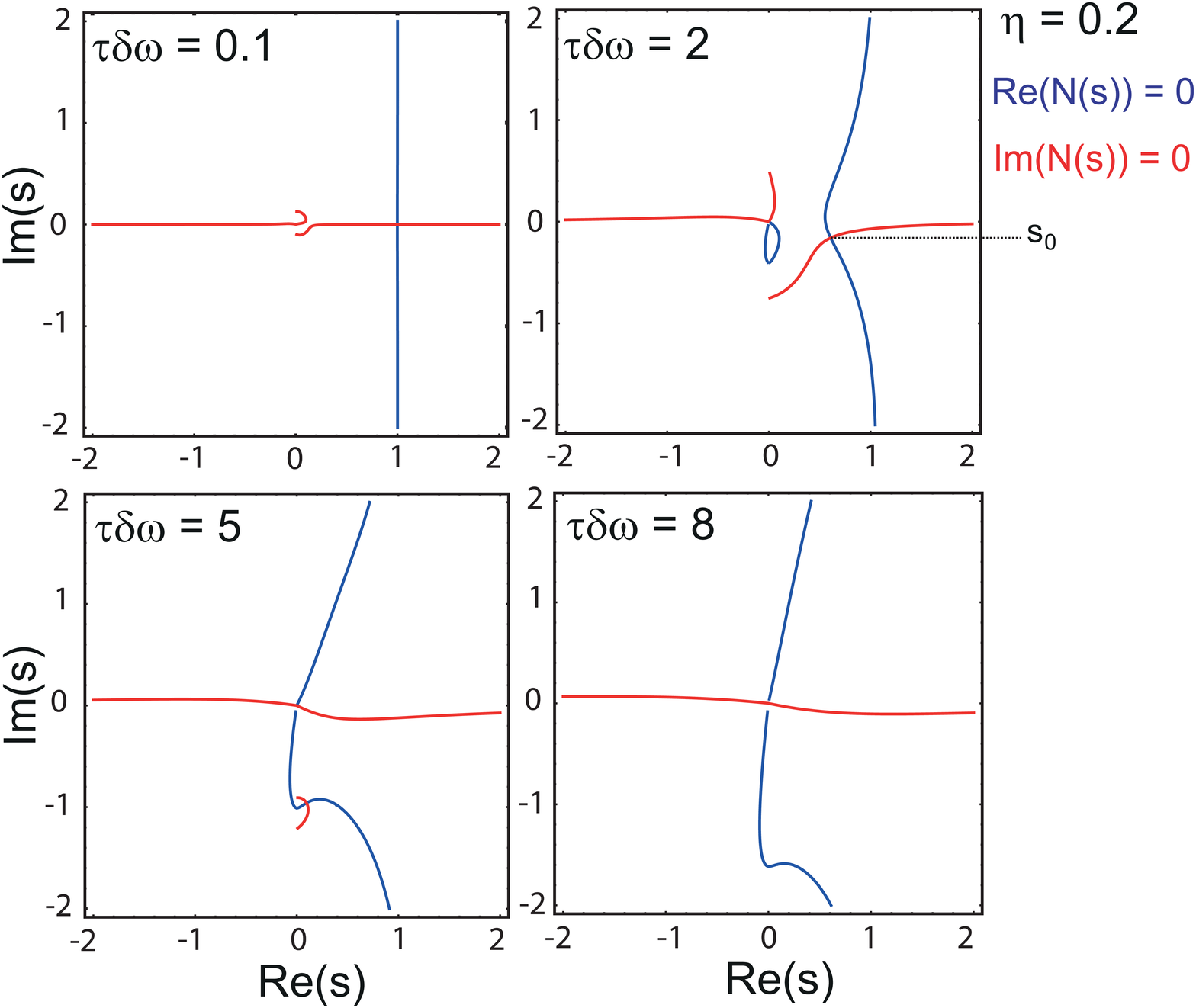}
\caption{
Kontur-Darstellung der Funktion $N(s)=0$ in der komplexen $s$-Ebene für unterschiedliche Werte des Parameters $\tau\delta\omega$ und für ein Volumenverhältnis $\eta=0,2$ mit $N(s)$ aus Gl. \eqref{eq:Nenner}. Die blaue Linie ($\text{Re}(N(s))=0$) und die rote Linie ($\text{Im}(N(s))=0$) bestehen aus Punkten für die der Realteil bzw. Imaginärteil von $N(s)$ verschwindet. Somit gibt der Schnittpunkt der beiden Linien die Position der Nullstelle $s_0$ des Nenners $N(s)$ in der komplexen Ebene an. Für das Volumenverhältnis $\eta =0,2$ ist der Grenzparameter $\taudom_0 \approx 7,29$ entsprechend Gl. \eqref{eq:TaudomBedingung}. Für $\taudom <\taudom_0$ tritt eine Nullstelle auf, für $\taudom > \taudom_0$ hingegen nicht mehr.}
\label{fig:contour}
\end{center}
\end{figure}

Hauptaufgabe dieser Arbeit ist es, den freien Induktionszerfall des oben beschriebenen Modells zu analysieren. Dafür wird die in Gl. \eqref{eq:Bromwich} beschriebene inverse Laplace-Transformation ausgeführt. Dazu ist es notwendig, die Singularität $s_0$ der Funktion $\hat{G}(s)$ zu finden. Diese stimmt mit der Nullstelle des Nenners $N(s)$ aus Gl. \eqref{eq:Nenner} überein:
\begin{equation}
\label{Def-s0}
N(s_0)=0 \,.
\end{equation}
Um die Nullstelle $s_0$ des Nenners $N(s)$ zu finden, wird die Funktion $N(s)=0$ in der komplexen Ebene betrachtet (siehe Abb. \ref{fig:contour}). Im Motional-Narrowing-Grenzfall $\tau\delta\omega \to 0$ ist die Singularität auf der reellen Achse bei $s_0=1$ lokalisiert. Für zunehmende Werte des Parameters $\taudom$ wandert die Singularität durch den vierten Quadranten (positiver Realteil und negativer Imaginärteil) in Richtung der imaginärer Achse.
Für den Wert $\tau\delta\omega = \tau\delta\omega_0$ ist die Nullstelle $s_0$ rein imaginär und dementsprechend auf der negativen, imaginären Achse lokalisiert. Für größere Werte des Parameters $\tau\delta\omega > \tau\delta\omega_0$ weist der Nenner $N(s)$ keine Nullstelle mehr auf. Empirisch kann beobachtet werden, dass $s_0$ an der imaginären Achse näherungsweise bei  
\begin{equation}
\label{eq:s0MaxApprox}
s_{0,\text{max}} \approx - \text{i}\eta \tau \delta \omega_0
\end{equation}
endet. Dementsprechend befindet sich die Nullstelle $s_0$ im Bereich der komplexen Ebene mit $0\leq\text{Re}(s_0) \leq 1$ auf der reellen Achse und zwischen  $0 \leq \text{Im}(s_0) \leq s_{0,\text{max}} \approx - \text{i} \eta \tau \delta \omega_0$ auf der imaginären Achse. Die Abhängigkeit des Realteils und des Imaginärteils der Nullstelle $s_0$ vom variablen Parameter $\taudom$ kann numerisch durch Lösen von $N(s_0)=0$ bestimmt werden und ist in Abb. \ref{fig:VerlaufNullstelle} visualisiert. 

\begin{figure}
\begin{center}
\includegraphics[width=9 cm]{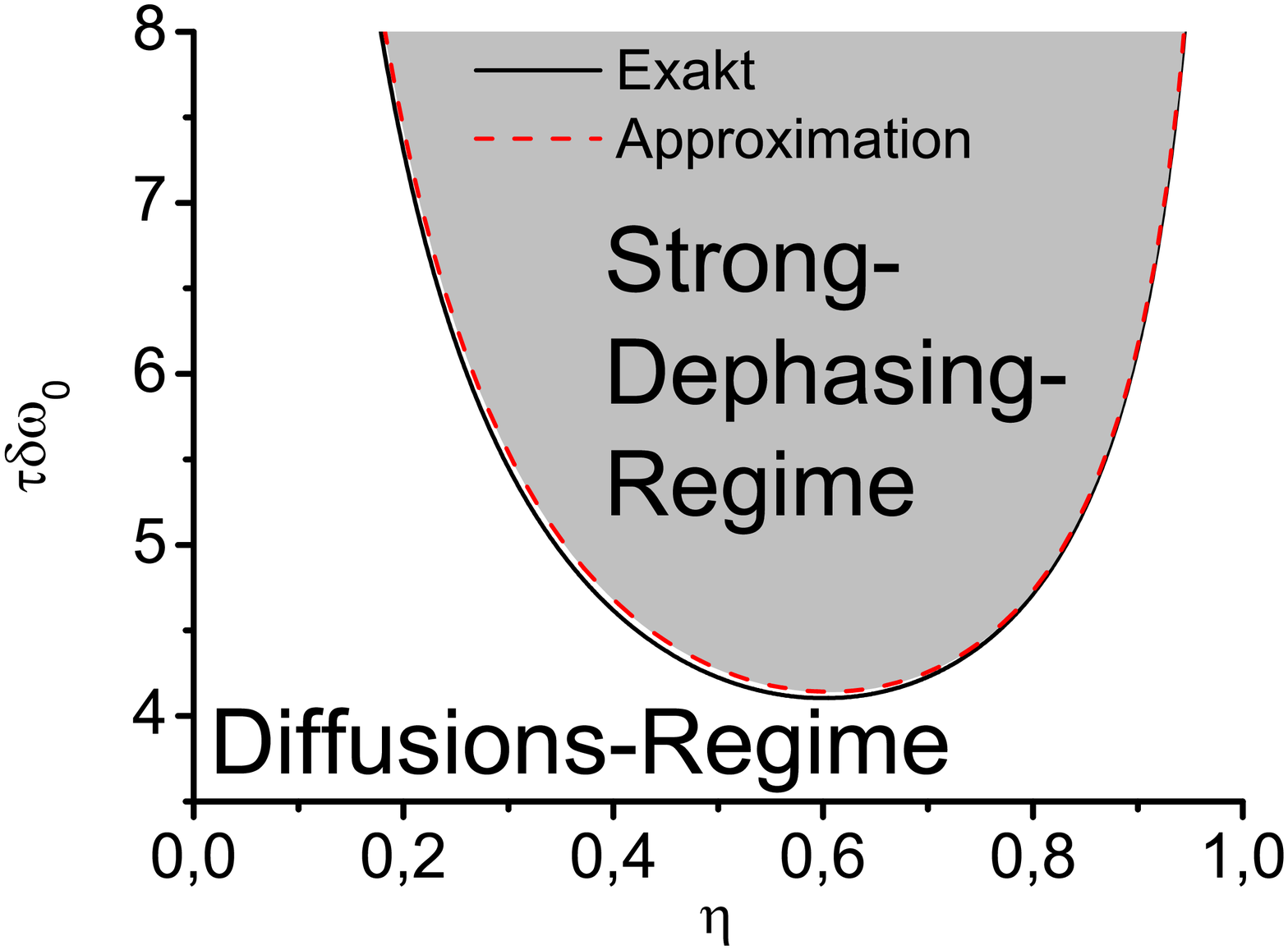}
\caption{
Klassifizierung des Diffusions-Regimes und des Strong-Dephasing-Regimes. Die schwarze Linie gibt die exakte numerische Lösung des Grenzparameters $\taudom_0$ an. Die rot gestrichelte Linie ist durch die Näherung aus Gl. \eqref{eq:TaudomBedingung} erhalten. Die Singularität $s_0$ von $\hat G(s)$ aus Gl. \eqref{eq:Gdach} existiert für $\taudom<\taudom_0$ und das zugehörige Bewegungsregime heißt Diffusions-Regime. Für $\taudom>\taudom_0$ tritt hingegen keine Singularität auf und das zugehörige Bewegungsregime nennt sich Strong-Dephasing-Regime.}
\label{fig:taudom0}
\end{center}
\end{figure}

Der spezifische Wert $\tau\delta\omega_0$, bei welchem die Singularität $s_0$ verschwindet, hängt einzig vom Volumenverhältnis $\eta$ ab. Durch die Näherung aus Gl. \eqref{eq:s0MaxApprox} kann der Grenzparameter $\taudom_0$ approximiert werden:
\begin{equation}
\label{eq:TaudomBedingung}
\taudom_0 \approx \frac{\pi}{9} \left[1+\frac{2}{\eta} \right] \sqrt{\frac{3}{1-\eta}} \,.
\end{equation}
In Abb. \ref{fig:taudom0} ist gezeigt, dass das Auftreten der Singularität $s_0$ eine Einteilung in verschiedene Diffusionsregime ermöglicht.
Der Parameter  $\tau\delta\omega_0$ zeigt eine U-förmige Abhängigkeit vom Volumenverhältnis (durchgezogene schwarze Linie in Abb. \ref{fig:taudom0}) mit einem Minimum bei $\eta=0,601$ und $\tau\delta\omega_0=4,1046$. Für $\taudom < \taudom_0$ ist das Diffusions-Regime das zugrundeliegende Regime. Das Strong-Dephasing-Regime ist durch die Bedingung $\taudom > \taudom_0$ charakterisiert. Die Näherung  für $\taudom_0$ aus Gl. \eqref{eq:TaudomBedingung} beschreibt den numerisch berechneten Wert sehr gut (siehe rote, gestrichelte Linie in Abb. \ref{fig:taudom0}) und lässt sich somit zur Bestimmung des Diffusionsregimes verwenden.

Neben der Singularität $s_0$ der Funktion $\hat{G}(s)$ aus Gl. \eqref{eq:Gdach} trägt auch die Integration um die Verzweigungslinie der Funktion $\hat{G}(s)$ zum freien Induktionszerfall bei. Die Verzweigungslinie ist auf der imaginären Achse im Bereich $-\taudom \leq \text{Im}(s) \leq +2\taudom$ lokalisiert (siehe Abb. \ref{fig:Bromwich}).

\subsection{Motional-Narrowing-Grenzfall}

Im Motional-Narrowing-Grenzfall ($\tau\delta\omega\to0$)  reduziert sich die Verzweigungslinie auf der imaginären Achse zu einer Integration um den Ursprung und verschwindet schließlich in diesem Grenzfall. Es trägt dann nur noch die Singularität $s_0=1$ bei. Im Motional-Narrowing-Grenzfall erfahren aufgrund der dominierenden Diffusion alle Spin-tragenden Teilchen im Mittel dieselbe lokale Larmor-Frequenz, sodass keine Dephasierung stattfindet. Wie erwartet, erhält man deshalb im Motional-Narrowing-Grenzfall durch Ausführen der inversen Laplace-Transformation eine konstante Magnetisierung $M(t)=1$. Experimentell findet ein Signalzerfall aufgrund von intrinsischer $T_2$-Relaxation der Form  $ \E^{-t/T_2}$ statt; dieser Umstand ist in dieser Arbeit zunächst vernachlässigt. Ebenso wird der Motional-Narrowing-Grenzfall erreicht, wenn das Volumenverhältnis $\eta$ gegen 0 strebt: $\lim_{\eta \to 0} p(\omega) = \delta(\omega)\,,$ mit der Diracschen Deltafunktion $\delta$ und der Frequenzverteilung $p(\omega)$ aus Gl. \eqref{eq:Fouriertransform}. In diesem Fall geht der äußere Radius $R_{\text{D}}$ gegen unendlich, was mit unbeschränkter Diffusion äquivalent ist, da die meisten Spins weit von der magnetisierten Kugel entfernt sind und dementsprechend der Einfluss der Frequenzverschiebung auf die diffundierenden Teilchen vernachlässigbar ist.

\subsection{Diffusions-Regime}

Mit abnehmendem Einfluss der Diffusion, d.h. wachsenden Werten von $\tau\delta\omega$, wandert die Singularität  Richtung imaginärer Achse wie in Abb. \ref{fig:contour} und später in Abb. \ref{fig:VerlaufNullstelle} gezeigt ist. Gleichzeitig vergrößert sich die Verzweigungslinie auf der imaginären Achse, siehe Abb. \ref{fig:Bromwich}. Im Diffusions-Regime sind die Diffusionseffekte auf die Magnetisierung immer noch stark und das Regime ist charakterisiert durch $0 < \taudom < \taudom_0$. Im biologischen Gewebe korrespondiert dieser Fall zu kleinen sphärischen, magnetischen Feldinhomogenitäten, um die Diffusion mit einer großen Diffusionskonstante $D$ und einer kleinen Frequenzverschiebung $\delta\omega$ stattfindet. Daher kann ein diffundierender Spin immer noch einen großen Bereich von unterschiedlichen Resonanzfrequenzen während der Dephasierung erfahren. Dadurch wird die aufgesammelte Phase gemittelt und der totale magnetische Zerfall findet langsam statt.

\begin{figure}
\begin{center}
\includegraphics[width=\textwidth]{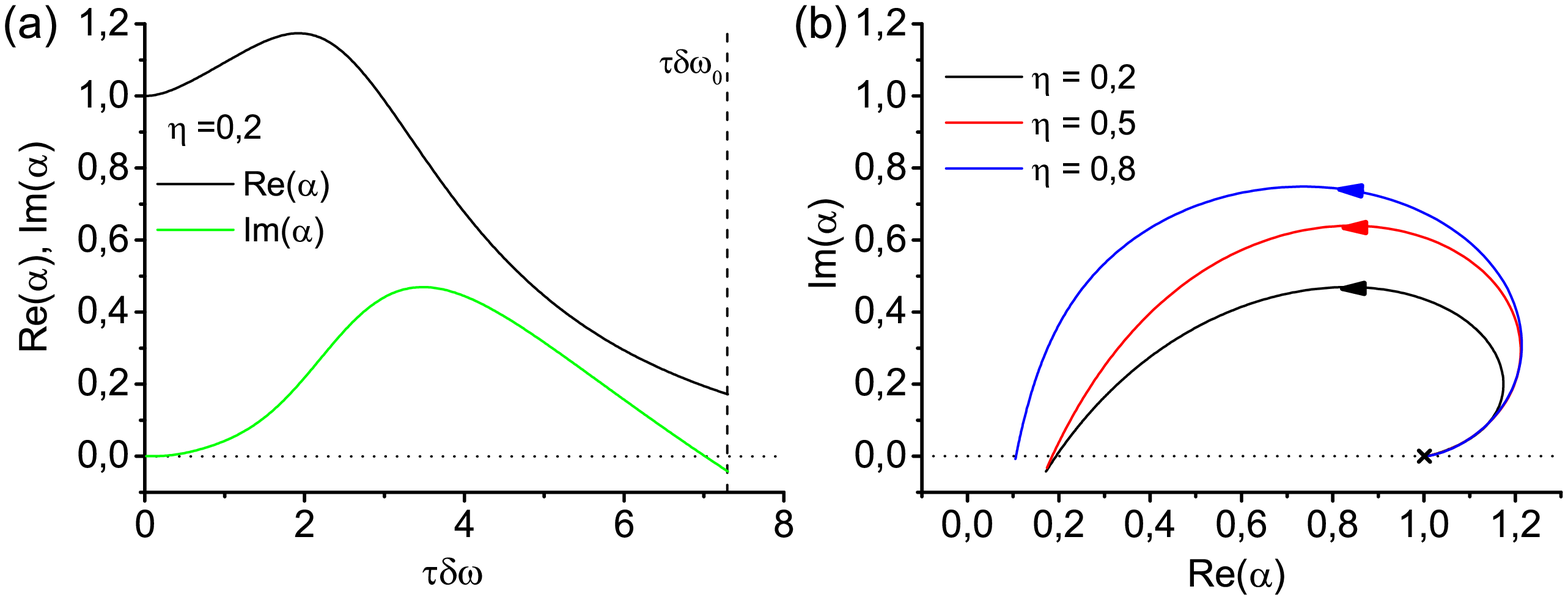}
\caption{
Amplitude $\alpha$ in Abhängigkeit von $\taudom$. (a) Realteil (schwarze, durchgezogene Linie) und Imaginärteil (grüne, durchgezogene Linie) der komplexen Amplitude $\alpha$ in Abhängigkeit von $\taudom$, wie in Gl. \eqref{eq:alpha} für ein Volumenverhältnis $\eta=0,2$ gegeben. Im Motional-Narrowing-Grenzfall $\tau\delta\omega\to0$ ist die Amplitude $\alpha$ rein reell mit dem Wert $\alpha=1$. Für wachsende Werte von $\taudom$ wird der Imaginärteil negativ, während der Realteil positiv bleibt. Da $\alpha$ eine Funktion der Nullstelle $s_0$ von $N(s)$ ist, verschwindet sie für $\tau\delta\omega>\taudom_0$, wie durch die schwarze, gestrichelte Linie markiert ist. (b) Parametrisierte Kurve der Amplitude $\alpha$ in der komplexen Ebene bei variablem Parameter $\taudom$ für verschiedene Volumenverhältnisse $\eta$. Alle Kurven starten im Motional-Narrowing-Grenzfall ($\tau\delta\omega=0$ wie durch das Kreuz in (b) bei $\alpha=1$ markiert ist) und setzen sich gegen den Uhrzeigersinn für wachsende Werte von $\taudom$ fort. Zudem wird die eingeschlossene Fläche der Kurve für größer werdende Volumenverhältnisse $\eta$ größer.}
\label{fig:alpha}
\end{center}
\end{figure}

Wie detailliert im Anhang \ref{AnhangC} beschrieben, trägt zur inversen Laplace-Transformation auch die Integration um die Verzweigungslinie der imaginären Achse bei. Letztlich kann die Zeitentwicklung der Magnetisierung wie folgt ausgedrückt werden:
\begin{equation} 
\label{eq:Result} 
M(t)=\E^{-\frac{t}{\tau}} \left[ \alpha \mathrm{e}^{s_0 \frac{t}{\tau}} + k(t) \right]
\end{equation}
mit der Amplitude
\begin{equation} 
\label{eq:alpha} 
\alpha = \frac{\scriptstyle{  \sqrt{3} [1-\eta] s_0^3}}{\scriptstyle{[1-\eta]s_0 \frac{3 s_0+1}{\sqrt{3}}+ \eta\sqrt{1-\frac{\mathrm{i} s_0}{\tau\delta\omega}} \left[\frac{s_0+\mathrm{i} \tau\delta\omega}{3}-\frac{[\tau\delta\omega]^2}{s_0+\mathrm{i} \tau\delta\omega}\right] \mathrm{arccoth}\left(\sqrt{\frac{1}{3}-\frac{\mathrm{i} s_0}{3 \tau\delta\omega}}\right)-\sqrt{1-\frac{\mathrm{i} s_0}{\eta  \tau\delta\omega}}\left[\frac{s_0+\mathrm{i}\eta\tau\delta\omega}{3}-\frac{ [\eta\tau\delta\omega]^2}{s_0+\mathrm{i} \eta  \tau\delta\omega}\right] \mathrm{arccoth}\left(\sqrt{\frac{1}{3}-\frac{\mathrm{i} s_0}{3 \eta\tau\delta\omega}}\right)}} \,,
\end{equation}
wobei die Abhängigkeit der komplexwertigen Amplitude $\alpha$ von dem Parameter $\taudom$ in Abb. \ref{fig:alpha} (a) für $\eta = 0,2$ dargestellt ist. 
Im Motional-Narrowing-Grenzfall ($\taudom \to 0$) wird die Amplitude $\alpha$ rein reell mit dem Wert $\alpha =1$. Zudem verschwindet die Amplitude $\alpha$ für $\taudom>\taudom_0$, da sie von der Nullstelle $s_0$ von $N(s)$ abhängt. Ein parametrisierter Plot der Amplitude $\alpha$ gegenüber $\taudom$ ist in Abb. \ref{fig:alpha} (b) für unterschiedliche Werte von $\eta$ gezeigt: Alle Kurven mit Startwert $\taudom=0$ beginnen bei $\alpha=1$ und setzen sich gegen den Uhrzeigersinn fort solange $\taudom < \taudom_0$ ist , danach verschwinden die Kurven.
 
\begin{figure}
\begin{center}
\includegraphics[width=\textwidth]{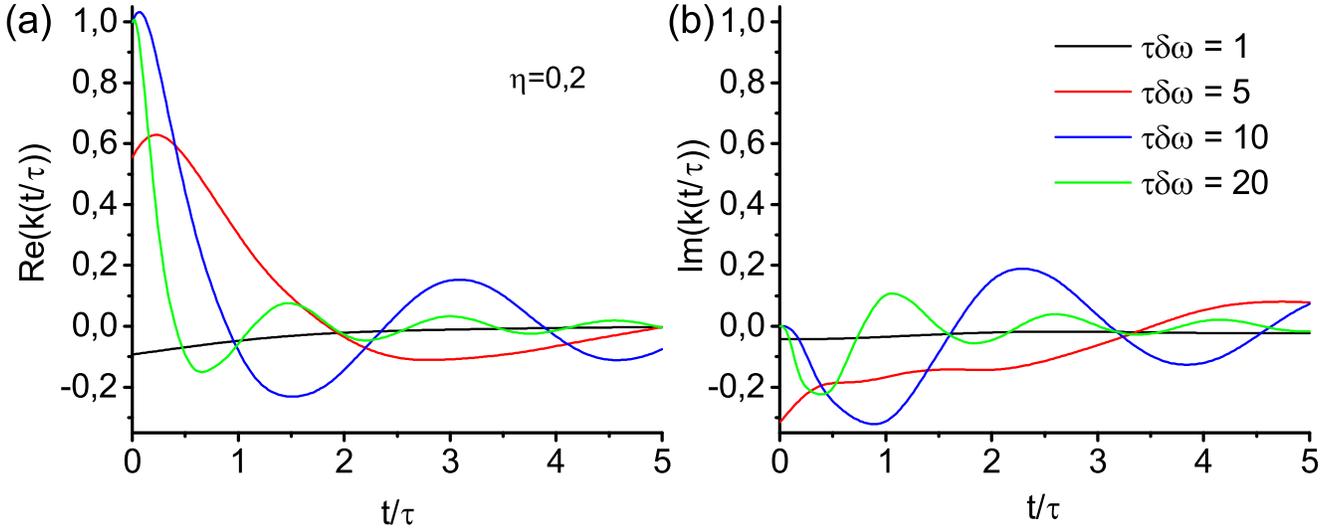}
\caption{
Zeitabhängigkeit der Funktion $k(t)$ für verschiedene Werte des Parameters $\taudom$ und einem Volumenverhältnis $\eta = 0,2$. (a) Realteil und (b) Imaginärteil der Funktion $k(t)$ für verschiedene Werte von $\taudom$, die charakteristisch für die zugehörigen Diffusionsregime sind. Im Diffusions-Regime (schwarze und rote Linien) ist der Startwert $k(0)$ durch $k(0) = 1-\alpha$ mit  $\alpha$ aus Gl. \eqref{eq:alpha} gegeben, während  man im Strong-Dephasing-Regime (blaue und grüne Linien) $k(0)=1$ findet.}
\label{fig:kt}
\end{center}
\end{figure}

Wie genauer in Anhang \ref{AnhangC} beschrieben, ist die Funktion $k(t)$ gegeben durch
\begin{align} 
\label{eq:k} 
k(t) &= \int\limits^{1}_{\eta}\!\!\frac{\scriptstyle{3\left[\taudom\right]^2 \left[1-\eta\right] \eta \left[2+x\right] \sqrt{\frac{1-x}{3}} \E ^{-\I x\delta\omega t}  \D x}}
{\scriptstyle{\pi^2\eta^2 \left[1+\frac{2}{x}\right]^2 \frac{x-1}{3}-\left[ 2\eta\left[ 1+\frac{2}{x}\right] \sqrt{\frac{1-x}{3}} \mathrm{arctanh} \left(\!\!\sqrt{\frac{1-x}{3}} \right)-2\left[1+\frac{2\eta}{x}\right] \sqrt{\frac{x}{3\eta}-\frac{1}{3}} \left[\frac{\pi}{2} - \mathrm{arctan}  \left(\!\!\sqrt{\frac{x}{3\eta}-\frac{1}{3}} \right) \right] -\left[3\I x \taudom+1\right] \left[1-\eta\right]\right]^2}}\\
\nonumber
&+\!\!\int\limits^{1}_{\eta}\!\!\frac{\scriptstyle{12 \left[\taudom\right]^2  \left[1-\eta\right] \eta \left[x-1\right] \sqrt{\frac{1+2x}{3}} \E ^{2\I x \delta\omega t} \D x}}
{\scriptstyle{\pi^2 \eta^2 \left[1-\frac{1}{x}\right]^2 \frac{1+2x}{3} +\left[2\eta\left[ 1-\frac{1}{x}\right] \sqrt{\frac{1}{3}+\frac{2x}{3}} \mathrm{arctanh} \left(\!\!\sqrt{\frac{1}{3}+\frac{2x}{3}} \right) -2\left[ 1-\frac{\eta}{x} \right] \sqrt{\frac{1}{3}+\frac{2x}{3\eta}}\mathrm{arccoth}\left(\!\!\sqrt{\frac{1}{3}+\frac{2x}{3\eta}} \right)+ \left[6\I x \taudom-1\right] \left[1-\eta\right]\right]^2}}\\
\nonumber 
&+\!\!\int\limits^{2}_{-1}\!\!\frac{\scriptstyle{3\left[\taudom\right]^2 \eta^2 \left[1-\eta \right] \left[\left[2-x \right] \sqrt{\frac{1+x}{3}} -\left[ 2-\eta x\right] \sqrt{\frac{1+\eta x}{3}}\right] \E ^{\I x \eta \delta\omega t} \D x}}{\scriptstyle{\pi^2 \left[ \left[\eta-\frac{2}{x}\right] \sqrt{\frac{1+\eta x}{3}}-\left[1-\frac{2}{x} \right] \sqrt{\frac{1+x}{3}}\right]^2+\left[2\left[\eta-\frac{2}{x}\right]\sqrt{\frac{1+\eta x}{3}}\mathrm{arctanh}\left(\!\!\sqrt{\frac{1+\eta x}{3}}\right)
-2\left[ 1-\frac{2}{x} \right] \sqrt{\frac{1+x}{3}}\mathrm{arctanh}\left(\!\!\sqrt{\frac{1+x}{3}}\right)+\left[3\I\taudom \eta x -1\right] \left[1-\eta\right]\right]^2}} \,
\end{align}
und in Abb. \ref{fig:kt} in Abhängigkeit der normalisierten Zeit $t/\tau$ dargestellt.
Aufgrund des Anfangswerts von $M(t=0)=1$ nimmt der Startwert der Funktion $k(t)$ im Diffusions-Regime den Wert $k(0)=1-\alpha$ an (siehe schwarze und rote Linien in Abb. \ref{fig:kt}). 
Des Weiteren sind die transversalen Komponenten sowie der Absolutwert und das Argument der Zeitentwicklung der Magnetisierung $M(t)$ in Abb. \ref{fig:signal} dargestellt.

Da die transversalen Komponenten der Magnetisierung zu einer komplexen Größe zusammengefasst werden, ist es sinnvoll ihre Zeitentwicklung in der komplexen Ebene darzustellen (siehe Abb. \ref{fig:Spiral}).

Offenbar zeigen die Komponenten der Magnetisierung oszillierendes Verhalten, was im Gegensatz zu einem einfachen monoexponentiellen Zerfall steht. In Analogie zum zylindrischen Fall erfüllt auch hier die Magnetisierung die Differentialgleichung eines gedämpften, getriebenen, harmonischen Oszillators in der Form
\begin{equation}
\label{Oszi-DGL}
\left[ \tau^2 \frac{\mathrm{d}^2}{\mathrm{d} t^2} + \tau\frac{\mathrm{d}}{\mathrm{d}t} + s_0 [1-s_0] \right] M(t)= \mathrm{e}^{-\frac{t}{\tau}} l(t)
\end{equation}
mit dem Inhomogenitätsterm
\begin{equation}
\label{l}
l(t)=\left[ \tau^2 \frac{\mathrm{d}^2}{\mathrm{d} t^2} - \tau\frac{\mathrm{d}}{\mathrm{d}t} + s_0 [1-s_0] \right] k(t)\,.
\end{equation}
Diese Gleichung lässt sich leicht durch Einsetzen von Gl. \eqref{eq:Result} verifizieren.
Zu beachten ist, dass sowohl der treibende Term als auch der Dämpfungsterm komplexe Größen sind, was zur Existenz der beiden transversalen Komponenten $M_x(t)$ und $M_y(t)$ von $M(t)=M_x(t)+\text{i}M_y(t)$ korrespondiert. Das Problem lässt sich somit als System von zwei gekoppelten, getriebenen, gedämpften, harmonischen Oszillatoren beschreiben.

\begin{figure}
\begin{center}
\includegraphics[width=9 cm]{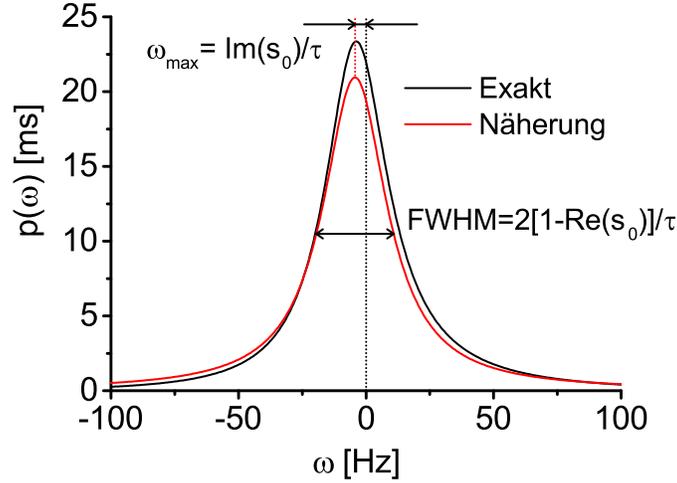}
\caption{
Exakte und genäherte Frequenzverteilung im Diffusions-Regime. Die exakte Frequenzverteilung nach Gl. \eqref{eq:Fouriertransform} und ihre Näherung nach Gl. \eqref{ResultPom-1} sind für ein Volumenverhältnis von $\eta=0,2$, Diffusionskoeffizient $D=2 \, \mu \text{m}^2/\text{ms}$, Radius $R=10 \, \mu \text{m}$ und Frequenzverschiebung  $\delta\omega = 100 \,  \text{Hz}$ dargestellt. Unter Benutzung von Gl. \eqref{eq:tau} ergibt sich die Korrelationszeit zu $\tau = 12,71 \, \text{ms}$. Die Nullstelle $s_0$ von $N(s)$ gegeben in Gl. \eqref{eq:Nenner} kann numerisch zu $s_0 \approx 0,81-0,05 \I$ bestimmt werden, was zu einem Maximum bei der Frequenz $\omega_{\text{max}} \approx - 4,32 \, \text{Hz}$ und einer Breite von $\text{FWHM}= 30,39 \, \text{Hz}$ führt.}
\label{fig:ApproxFS}
\end{center}
\end{figure}

\begin{figure}
\begin{center}
\includegraphics[width=\textwidth]{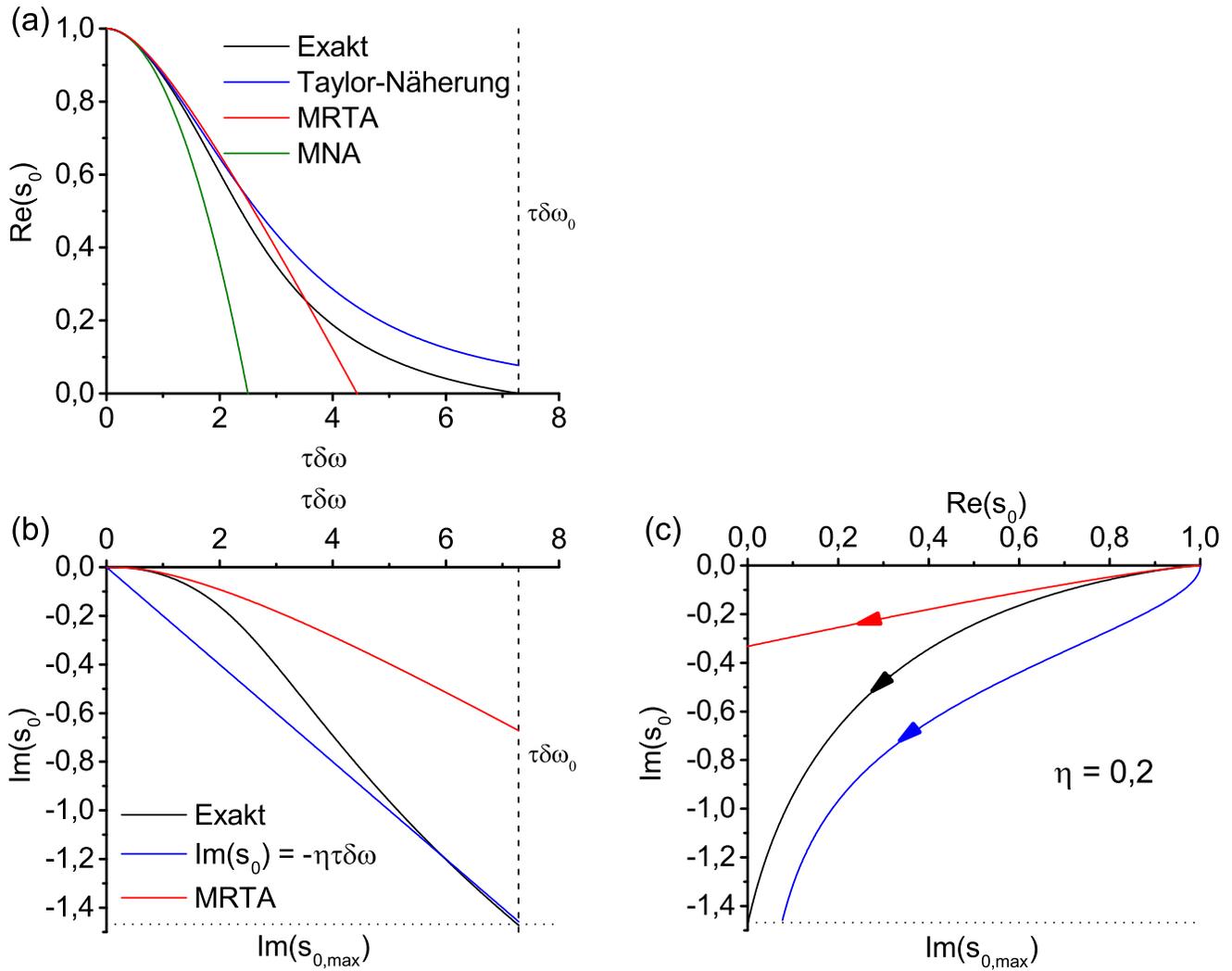}
\caption{
Singularität $s_0$ in Abhängigkeit des Parameters $\taudom$. (a) Realteil und (b) Imaginärteil der Singularität $s_0$ in Abhängigkeit von $\taudom$ für ein Volumenverhältnis $\eta=0,2$. Die exakte numerische Lösung der Gleichung $N(s_0)=0$ entsprechend Gl. \eqref{Def-s0} mit $N(s)$ gegeben in Gl. \eqref{eq:Nenner} ist in der schwarzen Linie gezeigt. Im Motional-Narrowing-Grenzfall $\tau\delta\omega = 0$ ist die Singularität bei $s_0=1$ positioniert. Mit wachsenden Werten von $\taudom$ fällt der Realteil der Nullstelle ab und erreicht $\text{Re}(s_0)=0$ für $\tau\delta\omega=\tau\delta\omega_0 \approx 7,29$, wie durch die vertikale, schwarz gestrichelte Linie in (a) und (b) gekennzeichnet ist. Der Imaginärteil weist in diesem Fall den Wert $\text{Im}(s_0)=\text{Im}(s_{0,\text{max}}) \approx -1,47 \approx -\eta \tau\delta\omega_0$ auf, wie mit der horizontalen, schwarzen, gepunkteten Linie in (b) und (c) markiert ist. In (a) und (b) sind die exakten numerischen Werte von $s_0$ in der schwarzen Linie mit der Mean-Relaxation-Time-Approximation (MRTA) aus Gl. \eqref{eq:s0-MRT} verglichen (siehe rote Linie). Der Realteil (a) der Singularität $s_0$ kann durch eine lineare Taylor-Entwicklung genähert werden (siehe blaue Linie), wie in Gl. \eqref{s0-Approx} gezeigt ist. Zudem kann man eine Motional-Narrowing-Approximation (MNA) durchführen (grüne Linie nach Gl. \eqref{eq:s0-MN}). Für Werte des Parameters $\taudom\approx\taudom_0$ kann der Imaginärteil der Nullstelle durch $\text{Im}(s_0) \approx - \eta\taudom$ genähert werden, wie in der blauen Linie in (b) gezeigt. In der schwarzen durchgezogenen Linie in (c) ist der numerisch exakte Pfad der Singularität im vierten Quadranten der komplexen Ebene für wachsende Werte des Parameters $\tau\delta\omega$ aufgezeigt. Dieser Pfad kann durch die Ergebnisse der MRTA-Näherung (rote Linie in (c)), erhalten in Gl. \eqref{eq:s0-MRT}, oder unter Benutzung der Taylor-Näherung für den Realteil (siehe Gl. \eqref{s0-Approx}) und einer linearen Näherung $\text{Im}(s_0) \approx - \eta\taudom$ für den Imaginärteil (blaue Linie in (c)) approximiert werden.}
\label{fig:VerlaufNullstelle}
\end{center}
\end{figure}

Für kleine Werte von $\taudom$ kann die Magnetisierung genähert werden, indem nur das Residuum der Singularität $s_0$ berücksichtigt wird und der Beitrag der Verzweigungslinie vernachlässigt wird. Dadurch wird die Magnetisierung als monoexponentieller Zerfall genähert:
\begin{align}\label{eq:Ms0}
M(t) & \approx \alpha \E^{[s_{0}-1] \frac{t}{\tau}}\\
\label{eq:MApprox}
& \approx \E^{[s_{0}-1] \frac{t}{\tau}} \,,
\end{align}
mit $\alpha\approx 1$ für kleine Werte des Parameters $\taudom$ wie in Abb.  \eqref{fig:alpha} (a) zu sehen ist. In Gl.\eqref{eq:MApprox} ist eine enge Verbindung der Singularität $s_0$ mit der Relaxationsrate des monoexponentiell genäherten Zerfalls zu erkennen. Mit dieser Näherung der Magnetisierung $M(t)$ und der Symmetrieeigenschaft $M(-t)=M^*(t)$ ist es möglich, die Frequenzverteilung gemäß Gl. \eqref{eq:RueckFT} zu berechnen:
\begin{align}
\label{ResultPom}
p(\omega) & \approx \frac{\tau}{\pi}\frac{\text{Re}\left(\alpha [1-s_0^*]\right)+\tau\omega\text{Im}(\alpha)}{\left[1-\text{Re}(s_0)\right]^2+\left[\tau\omega-\text{Im}(s_0)\right]^2} \\
\label{ResultPom-1}
& \approx \frac{\tau}{\pi}\frac{1-\text{Re}(s_0)}{\left[1-\text{Re}(s_0)\right]^2+\left[\tau\omega-\text{Im}(s_0)\right]^2} \,,
\end{align}
wobei der letzte Ausdruck direkt aus der Tatsache folgt, dass $\alpha \approx 1$ für kleine Werte des Parameters $\taudom$. Dieser Ausdruck der Frequenzverteilung ist eine passende Näherung der Frequenzverteilung in Gl. \eqref{eq:Fouriertransform} für kleine Werte von $\taudom$. Aufgrund der Lorentz-Form dieser Frequenzverteilung kann die Verschiebung der Frequenzverteilung gegenüber dem Ursprung und die Breite (FWHM) in Abhängigkeit der Singularität $s_0$ bestimmt werden. Beides sind wichtige physikalische Parameter, um den Einfluss der Diffusion auf die Frequenzverteilung zu beurteilen. Die Peak-Position $\omega_{\text{max}}$ erhält man in dieser Näherung bei $\omega_{\text{max}} \approx  \text{Im}(s_0)/\tau$ und die Halbwertsbreite (FWHM) beträgt $2[1-\text{Re}(s_0)]/\tau$. In Abb. \ref{fig:ApproxFS} ist die Frequenzverteilung aus Gl. \eqref{eq:Fouriertransform} mit der eben beschriebenen Näherung aus Gl. \eqref{ResultPom-1} verglichen.

Der aus Gl. \eqref{eq:MApprox} bekannte Zusammenhang zwischen der Zerfallsrate und der Lage der Singularität $s_0$ zeigt, dass das Verständnis des Verlaufs der Singularität $s_0$ in der komplexen Ebene in Abhängigkeit der Parameter $\eta$ und $\taudom$ von großer Bedeutung ist. Aus diesem Grund wird dieser Verlauf genauer untersucht:

Ausgehend vom Motional-Narrowing-Grenzfall kann eine Taylor-Entwicklung von $N(s)$ um den Punkt $s=1$ in erster Ordnung durchgeführt werden. Dies liefert für den Realteil den Zusammenhang:
\begin{equation}
\label{s0-Approx}
\scriptstyle{\text{Re}\left(s_0\right) \approx \text{Re}}\left(\!\frac{\scriptstyle{\sqrt{1-\frac{\I}{\taudom}} \left[8 \eta^2 [\taudom]^2-4\I\eta\taudom+1 \right] \text{arctanh}\left(\sqrt{\frac{3\eta\taudom}{\eta\taudom-\I}}\right)-\eta  \sqrt{1-\frac{\I}{\eta  \taudom}} \left[\eta  \left[8 [\taudom]^2-4 \I \taudom+1\right] \text{arctanh}\left(\sqrt{\frac{3\taudom}{\taudom-\I}}\right)-6 \I [1-\eta] \taudom \sqrt{\frac{1}{3}-\frac{\I}{3\taudom}} \right]}}{\scriptstyle{\sqrt{1-\frac{\I}{\taudom}} \left[4 \eta^2  [\taudom]^2- 2\I \eta\taudom -1\right] \text{arctanh}\left(\sqrt{\frac{3\eta\taudom}{\eta\taudom-\I}}\right)-\eta  \sqrt{1-\frac{\I}{\eta  \taudom}} \left[ \eta  \left[4[\taudom]^2-2 \I \taudom-1 \right] \text{arctanh}\left(\sqrt{\frac{3\taudom}{\taudom-\I}}\right) - 12 \I [1-\eta] \taudom \sqrt{\frac{1}{3}-\frac{\I}{3 \taudom}}\right]}} \!\right)\,.
\end{equation}
Für den Imaginärteil ist es günstig entsprechend Gl. \eqref{eq:s0MaxApprox} die Näherung $\text{Im}(s_0) \approx -\eta\tau\delta\omega$ zu benutzen.

Diese Näherung ist in Abb. \ref{fig:VerlaufNullstelle} (a) mit der numerischen Lösung von Gl. \eqref{Def-s0} für ein Volumenverhältnis von $\eta = 0,2$ verglichen (blaue Linie). Der Verlauf von $s_0$ in der komplexen Ebene ist in Abb. \ref{fig:VerlaufNullstelle} (c) gezeigt.

Für kleine Werte des Parameters $\taudom$ kann der monoexponentielle freie Induktionszerfall von Gl. \eqref{eq:R2} durch eine rein reelle Relaxationsrate $R_2^{'} = \tau \langle \omega^2(\mathbf{r})\rangle$ genähert werden, welche von der Korrelationszeit $\tau$ in Gl. \eqref{eq:tau} und dem Erwartungswert $\langle \omega^2(\mathbf{r})\rangle = \frac{1}{V}\int_V \omega^2(\mathbf{r}) \mathrm{d}^3 \mathbf{r}= \frac{4}{5} \eta\delta\omega^2$ abhängt \cite{Ziener05a}. Schlussendlich kann damit der freie Induktionszerfall wie folgt approximiert werden:
\begin{equation}
\label{eq:MNApprox}
M(t) \approx \E^{-\frac{4}{5}\eta\tau\delta\omega^2t}\,.
\end{equation}
Damit lässt sich durch Vergleichen mit Gl. \eqref{eq:MApprox} eine Näherung für die Nullstelle $s_0$ finden:
\begin{equation}
\label{eq:s0-MN}
s_0 \approx 1-\frac{4}{5}\eta\left[\tau\delta\omega\right]^2 \,.
\end{equation}
Diese Näherung ist in der grünen Linie in Abb. \ref{fig:VerlaufNullstelle} (a) dargestellt.

Eine andere Möglichkeit um eine Näherung der Singularität $s_0$ zu finden, ist es, die monoexponentielle Relaxationsrate in Gl. \eqref{R2strich-MRT} zu betrachten. Dies führt mittels Gl. \eqref{eq:MApprox} zu folgender Approximation:
\begin{equation}
\label{eq:s0-MRT}
s_0 \approx 2-\frac{1-\eta}{H\left(\frac{1}{\eta\taudom}\right)-\eta H\left(\frac{1}{\taudom}\right)} = 2- \frac{1-\eta}{1-\eta-N(1)}\,.
\end{equation}
Auch diese Näherung ist in Abb. \ref{fig:VerlaufNullstelle} dargestellt (rote Linie).

\subsection{Strong-Dephasing-Regime}
Mit Zunahme von $\taudom$ verschwindet die Singularität $s_0$ und dementsprechend auch der Beitrag der Amplitude $\alpha$ zur Magnetisierung $M(t)$. Dieses Regime, das sogenannte Strong-Dephasing-Regime, ist durch die Bedingung $\taudom > \taudom_0$ charakterisiert und die Dephasierung ist hauptsächlich durch die statische Frequenz $\delta\omega$ bestimmt. Anschaulich ist dies äquivalent zu Spin-Trajektorien, die im dreidimensionalen Dipolfeld nur einen kleinen Weg zurücklegen und somit auch nur einen kleinen Bereich von Frequenzen abdecken. Die Dephasierung ist somit stärker  als in den vorangegangenen Regimen.
Letztlich kann die Zeitentwicklung der Magnetisierung als 
\begin{equation}
\label{Mshort}
M(t)= \E^{-\frac{t}{\tau}} k(t)\,,
\end{equation}
geschrieben werden, wobei $k(t)$ in Gl. \eqref{eq:k} gegeben ist.

\begin{figure}
\begin{center}
\includegraphics[width=\textwidth]{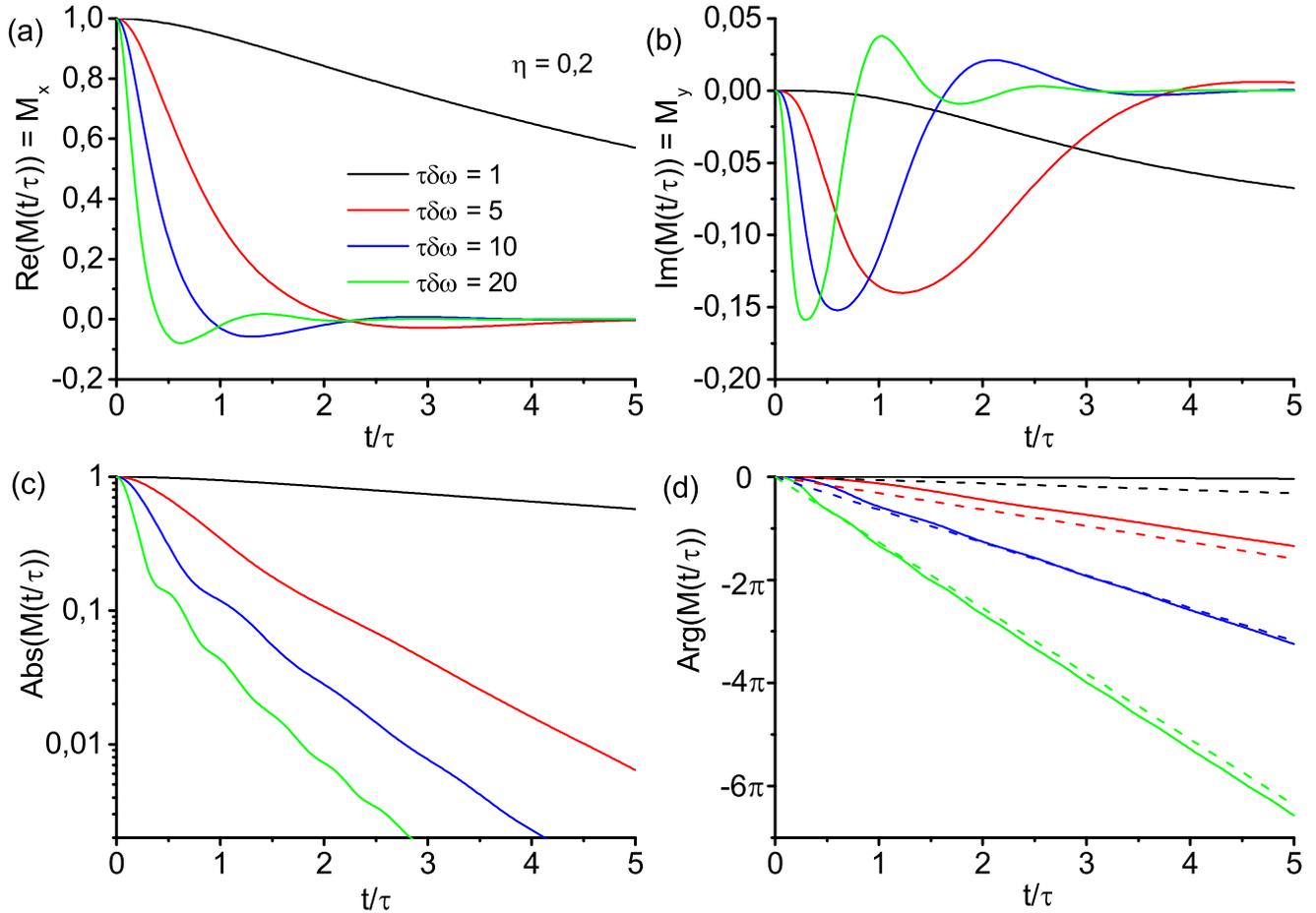}
\caption{
Zeitentwicklung der Magnetisierung $M(t)$ für ein Volumenverhältnis $\eta = 0,2$ und für verschiedene Werte von $\taudom$ erhalten aus Gl. \eqref{eq:Result} und in durchgezogenen Linien gezeichnet. (a) Realteil und (b) Imaginärteil sowie (c) Absolutwert und (d) Argument von $M(t)$ für verschiedene Werte von $\taudom$. Die gestrichelte Linie in (d) zeigt die Näherung $\text{Arg}(M(t))\approx - \eta \delta\omega t$ in Übereinstimmung mit Gl. \eqref{eq:MApprox} mit  $\text{Im}(s_0) \approx -\eta\tau\delta\omega$, wie in der blauen Linie in Abb. \ref{fig:VerlaufNullstelle} (b) dargestellt ist. Eindeutig zu erkennen ist, dass die Magnetisierung für größere Werte von $\taudom$ schneller abfällt und mit höheren Frequenzen oszilliert.}
\label{fig:signal}
\end{center}
\end{figure}

Die Zeitentwicklung der Magnetisierung im Strong-Dephasing-Regime ist durch die blaue und grüne Kurve in Abb. \ref{fig:signal} repräsentiert. Zudem ist die Zeitabhängigkeit der Funktion $k(t)$ in Abb. \ref{fig:kt} dargestellt (blaue und grüne Linie). Aufgrund des Anfangswerts der Magnetisierung $M(t=0)=1$ gilt in diesem Regime $k(t=0)=1$, wie ebenfalls in Abb. \ref{fig:kt} zu erkennen ist.

\subsection{Static-Dephasing-Grenzfall}
Im Grenzfall $\taudom \to \infty$ kann der Integrand der Funktion $k(t)$ in Gl. \eqref{eq:k} erheblich vereinfacht werden und die Static-Dephasing-Magnetisierung (Index $0$) von Gl. \eqref{Mshort} vereinfacht sich zu
\begin{equation}
\label{M02}
M_0(t)= \frac{1}{1-\eta} \frac{1}{3\sqrt{3}} \int\limits_{-1}^{+2} \mathrm{d}y \frac{\mathrm{e}^{\mathrm{i}y\delta\omega t}}{y^2}\left[ \eta [2-y] \sqrt{1+y}-\Theta(y+\eta)\Theta(2\eta-y)[2\eta-y]\sqrt{1+\frac{y}{\eta}} \right]\,,
\end{equation}
mit der Heaviside-Sprungfunktion $\Theta$. In diesem Grenzfall werden die Diffusionseffekte vernachlässigt ($D=0$). Der Ausdruck $M_0(t)$ aus der letzten Gleichung stimmt mit Gl. \eqref{eq:M0Numerik} und Gl. \eqref{M01} überein. Zudem kann dieses Ergebnis durch die Fourier-Transformation (siehe Gl.  \eqref{FT}) der Frequenzverteilung des Static-Dephasing-Regime $p_0(\omega)$ erhalten werden, welche in Gl. \eqref{eq:p0} des Anhangs \ref{AnhangA} angegeben ist. Für eine numerische Berechnung des Integrals in Gl. \eqref{M02} empfiehlt es sich, zuerst den Integranden einschließlich der Heaviside-Funktion zu berechnen, um damit einen konvergierenden Integrand zu erhalten. Im Static-Dephasing-Grenzfall folgt die Symmetrieeigenschaft $M_0(-t) = M_0^{*}(+t)$ direkt aus Gl. \eqref{M02}. Diese Eigenschaft kann jedoch auf alle anderen Diffusionsregime für $D>0$ verallgemeinert werden.
\begin{figure}
\begin{center}
\includegraphics[width=9 cm]{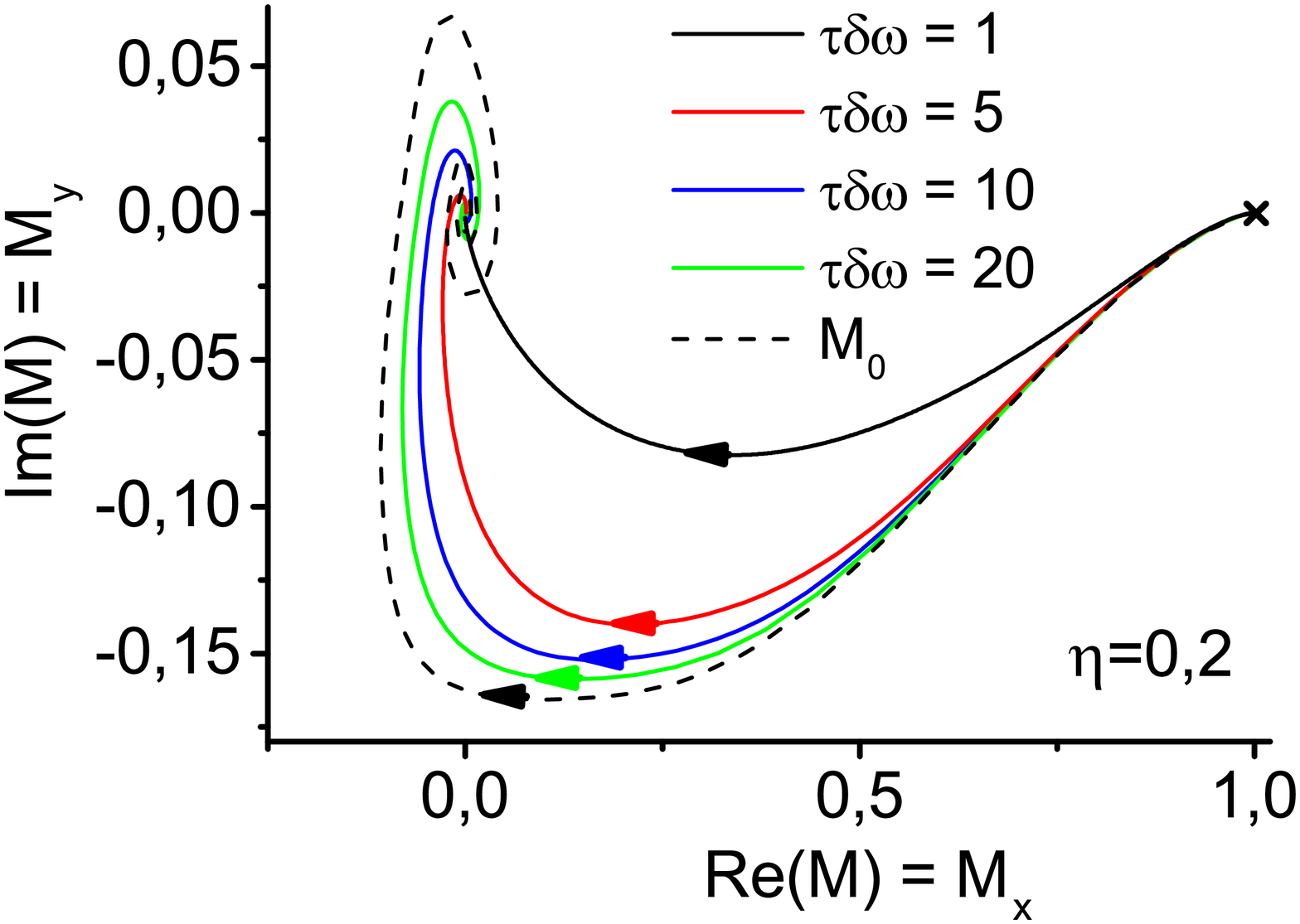}
\caption{Zeit-parametrisierte Magnetisierung $M(t)$ in der komplexen $M$-Ebene für $\eta=0,2$ und variierenden Werten von $\taudom$ wie in Gl. \eqref{eq:Result} gegeben. Im Motional-Narrowing-Grenzfall nimmt die Magnetisierung den konstanten Wert $M_x=1$ und $M_y=0$ für alle Zeiten $t$ an, wie durch das Kreuz dargestellt wird. Alle Kurven starten bei  $M_x=1$ und $M_y=0$ und konvergieren gegen den Ursprung. Die einhüllende Kurve der Static-Dephasing-Magnetisierung $M_0$ ist durch Gl. \eqref{M02} erhalten, die identisch zu Gl. \eqref{eq:M0Numerik} und Gl. \eqref{M01} ist; sie weist die größte Amplitude auf. Der Verlauf der Magnetisierung in der komplexen Ebene zeigt spiralförmiges Verhalten. Im Static-Dephasing-Grenzfall kann dieses Verhalten auf Cornu-Spiralen und Nielsen-Spiralen zurückgeführt werden (siehe Anhang \ref{AnhangA}).}
\label{fig:Spiral}
\end{center}
\end{figure}

\section{Experimentelle Verifikation}
Um die Anwendbarkeit der Strong-Collison-Näherung und der geometrischen Annahmen über das Gewebe zu überprüfen, eignet sich beispielsweise die Analyse des freien Induktionszerfalls an Lungengewebe. Dort erzeugt der Suszeptibilitätsunterschied zwischen luftgefüllten Alveolen und umliegenden Gewebe ein dreidimensionales, magnetisches Dipolfeld \cite{Cutillo96}. In der bisherigen Analyse wurde der Einfluss der intrinsischen $T_2$-Relaxation vernachlässigt. Unter Berücksichtigung dieses zusätzlichen Signalzerfalls ergibt sich das Gesamtsignal $S(t)$ als:
\begin{equation}
\label{eq:SignalGes}
S(t) = M(t) \E^{-\frac{t}{T_2}} \,.
\end{equation}
Analog zu Gl. \eqref{eq:RueckFT} ist das Frequenzspektrum $\sigma(\omega)$ des Gesamtsignals $S(t)$ gegeben durch
\begin{equation}
\label{FrequenzSpekGes}
\sigma(\omega)= \frac{1}{2\pi} \int \limits_{-\infty}^{+\infty} \D t  S(t) \E^{-\I \omega t}\,.
\end{equation}
Unter Verwendung von Gl. \eqref{eq:SignalGes} und dem Faltungssatz lässt sich dieser Ausdruck wie folgt vereinfachen:
\begin{equation}
\sigma(\omega) = \frac{1}{\pi} \int \limits_{-\infty}^{+\infty} \D \omega' p(\omega') \frac{T_2}{1+\left[T_2 \left[\omega-\omega' \right]\right]^2}\,,
\label{eq:Faltung}
\end{equation}
mit der Frequenzverteilung $p(\omega')$ aus Gl. \eqref{eq:Fouriertransform} und der Lorentz-Kurve $T_2/[1+[T_2 [\omega-\omega']]^2]$, welche die intrinsische $T_2$-Relaxation berücksichtigt. Das gemessene Frequenzspektrum $\sigma(\omega)$ ist somit eine Faltung aus der berechneten Frequenzverteilung $p(\omega)$ und der Lorentz-Kurve $T_2/[1+[T_2 \omega]^2]$. Die Breite dieser Lorentz-Kurve wird durch die intrinsische $T_2$-Relaxation des Lungengewebes bestimmt. Im Static-Dephasing-Grenzfall kann das Integral aus Gl. \eqref{eq:Faltung} analytisch gelöst werden (siehe Gl. \eqref{eq:FaltungSD} in Anhang \ref{AnhangA}).

Mittels Gl. \eqref{eq:Faltung} können nun die theoretischen Vorhersagen mit Lungenspektren verglichen werden, die von Mulkern \emph{et al.} gemessen wurden (siehe Abb. 3 in \cite{Mulkern14}). Die intrinsische $T_2$-Relaxationszeit des Lungengewebes beträgt $T_2=40 \text{ ms}$ \cite{Wild12}.
\begin{figure}[h]
\begin{center}
\includegraphics[width=9cm]{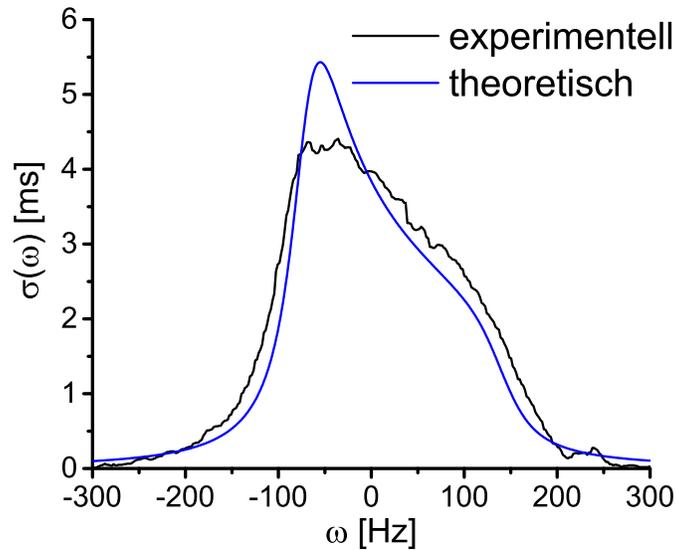}
\caption{
Experimentelles und theoretisches Frequenzspektrum der Lunge. Das experimentelle Frequenzspektrum wurde mit einer PRESS-Sequenz \cite{Bottomley84, Bottomley87} bei 1,5 T an einem gesunden Freiwilligen von Mulkern \emph{et al.} gemessen (siehe Abb. 3 in \cite{Mulkern14}). Das theoretische Spektrum wurde nach Gl. \eqref{eq:Faltung} mit $p(\omega')$ aus Gl. \eqref{eq:Fouriertransform} für realistische Werte des Volumenverhältnis $\eta=0,8$, des Alveolarradius $R=150\, \mu \text{m}$ und des Diffusionskoeffizienten $D= 1\, \mu \text{m}^2\text{ms}^{-1}$ dargestellt \cite{Cutillo96}. Damit ergibt sich eine Korrelationszeit von $\tau = 4,0 \text{ s}$ und der Fit der theoretischen Kurve an die Daten liefert eine Frequenzverschiebung von $\delta\omega = 79 \text{ Hz}$. Entsprechend Abb. \ref{fig:taudom0} ist somit das zugrundeliegende Regime das Strong-Dephasing-Regime. Das theoretische Spektrum $\sigma(\omega)$ ergibt sich aus der Faltung des Frequenzspektrum $p(\omega)$ aus Gl. \eqref{eq:Fouriertransform}, in dem keine $T_2$-Effekte berücksichtigt werden, mit einer Lorentz-Kurve, die von $T_2$ abhängt (siehe Gl.  \eqref{eq:Faltung}). Entsprechend \cite{Wild12} wird für das vorliegende Lungengewebe $T_2 = 40 \text{ ms}$ angenommen.}
\label{fig:Mulkern}
\end{center}
\end{figure}
In Abb. \ref{fig:Mulkern} ist das gemessene Frequenzspektrum dargestellt (schwarze Linie) und mit dem theoretischen Spektrum $\sigma(\omega)$ für realistische Werte der Parameter $\eta = 0,8\text{; } D=1\, \mu \text{m}^2\text{ms}^{-1} \text{und } R = 150\, \mu \text{m}$ verglichen (blaue Linie). Aus diesen Parametern kann die Korrelationszeit mittels Gl. \eqref{eq:tau} zu $\tau = 4,0 \text{ s}$ bestimmt werden. Die Frequenzverschiebung ist durch den Fit der theoretischen Kurve an die experimentellen Daten in Abb. \ref{fig:Mulkern} auf $\delta\omega =79 \text{ Hz}$ festgelegt. Entsprechend Gl. \eqref{eq:TaudomBedingung} beträgt der Grenzparameter $\taudom_0 = 4,7$ für das Volumenverhältnis $\eta= 0,8$. Aufgrund der angenommenen Daten ergibt sich damit $\taudom_0 = 4,7 < \taudom = 316$. Somit ist das zugrundeliegende Regime das Strong-Dephasing-Regime (siehe auch Abb. \ref{fig:taudom0}).

Wie in Abb. \ref{fig:Mulkern} zu erkennen ist, weisen das experimentelle und das theoretische Frequenzspektrum ähnliche Verläufe insbesondere an den Flanken des Peaks auf. Beide Peaks sind asymmetrisch und ihre Breiten sind ähnlich groß.

\chapter{Diskussion}
In der vorliegenden Arbeit wird die Zeitentwicklung der transversalen Magnetisierung, die von der Dephasierung von Spin-tragenden Teilchen im lokalen Dipolfeld eines sphärischen Objektes bestimmt ist, unter Beachtung von Diffusionseffekten analysiert. Ein detailliertes Verständnis des Zerfalls der Magnetisierung erlaubt es, die Signalentstehung z.B. in Lungengewebe zu quantifizieren. Dort können Alveolen als sphärische, magnetische Feldinhomogenitäten aufgefasst werden, da der Suszeptibilitätsunterschied zwischen der Luft in den Alveolen und dem umgebenden Gewebe ein dreidimensionales Dipolfeld generiert. Dadurch tritt im umgebenden Gewebe der Alveolen Spindephasierung auf \cite{Cutillo96}. Auf ähnliche Weise kann die Signalentstehung in Geweben mit kleinen Eisenoxid-Teilchen (USPIO's$=$ Ultrasmall Superparamagnetic Iron Oxides) oder Blutrückständen und Eisenansammlungen bei neurodegenerativen Erkrankungen präziser studiert werden \cite{Sedlacik13,Bendszus03}.

Im Fall von dominanter Diffusion, dem Motional-Narrowing-Grenzfall zerfällt die Magnetisierung monoexponentiell mit nur einer transversalen Komponente. Mit abnehmendem Einfluss der Diffusion und dementsprechend einem Übergang zum Static-Dephasing-Regime, findet eine Zunahme beider transversaler Komponenten $M_x$ und $M_y$ der Magnetisierung statt. Sie zerfallen nicht monoexponentiell, sondern zeigen auch oszillierendes Verhalten.

Der Diffusionsprozess wird mathematisch durch die Strong-Collision-Näherung beschrieben, wodurch ein analytischer Ausdruck der Laplace-Transformierten der Magnetisierung in Gl. \eqref{eq:LT} angegeben werden kann. Eine detaillierte Analyse der inversen Laplace-Transformation in der komplexen Ebene erlaubt es verschiedene Diffusionsregime zu unterscheiden: Aufgrund der mathematischen Form der inversen Laplace-Transformation des freien Induktionszerfalls existiert eine Singularität $s_0$ (siehe Gl. \eqref{eq:Gdach}), die im Motional-Narrowing-Regime auf der reellen Achse bei $s_0=1$ startet und mit zunehmenden Werten von $\taudom$ Richtung imaginärer Achse wandert, siehe Abb. \ref{fig:VerlaufNullstelle}. Der Weg der Singularität $s_0$ durch die komplexe Ebene definiert das Diffusionsregime. Für unbeschränkte Diffusion ($\taudom \to \infty$) liegt der Motional-Narrowing-Grenzfall vor, in dem das Signal nur aufgrund von intrinsischer $T_2$-Relaxation zerfällt. Die Singularität liegt bei $s_0=1$ auf der reellen Achse. Mit abnehmendem Einfluss der Diffusion ($\taudom < \taudom_0$) zerfällt das Signal schneller und oszilliert stärker (siehe Abb. \ref{fig:signal}); dieser Fall ist als Diffusions-Regime bekannt. Die Singularität $s_0$ wandert Richtung negativer imaginärer Achse und erreicht diese für $\taudom=\taudom_0$. Für größere Werte von $\taudom$ ist die Singularität nicht mehr vorhanden, dieser Fall ist als Strong-Dephasing-Regime klassifiziert. Er setzt sich in den Grenzfall des Static-Dephasing-Regime für $\tau\delta\omega\to\infty$ fort, in dem die Diffusion vernachlässigt wird ($D=0$).

Analog zum Dephasierungsprozess im zweidimensionalen Dipolfeld  verursacht durch ein zylindrisches Objekt, kann auch beim hier untersuchten dreidimensionalen Dipolfeld eine Analogie zum harmonischen Oszillator  aufgezeigt werden: Der Zerfall der Magnetisierung verhält sich wie ein gekoppelter, gedämpfter und getriebener harmonischer Oszillator (siehe Gl. \eqref{Oszi-DGL}). Im Unterschied zum zweidimensionalen Fall sind hier jedoch beide transversalen Komponenten der Magnetisierung relevant. Die Amplitude des dreidimensionalen Falls ist nur im Motional-Narrowing-Regime rein reell, für geringer werdenden Einfluss der Diffusion ist der Imaginärteil jedoch nicht vernachlässigbar. Zudem weist die Funktion $k(t)$ (die der Funktion $h(t)$ in Abb. 5 in \cite{Ziener12a} im zweidimensionalen Fall entspricht) einen imaginären Anteil auf, der in ähnlicher Weise wie der Realteil oszilliert. Die Oszillationen sind etwas schneller verglichen mit der Oszillation im zweidimensionalen Fall (z.B. für den Fall von $\taudom=5$). Der Absolutwert der Funktion $k(t)$ kann größere Werte als Eins annehmen (siehe Abb. \ref{fig:kt} für $\taudom=10$), da der exponentielle Faktor $\E^{-\frac{t}{\tau}}$ sicherstellt, dass $\vert M(t)\vert \leq 1$.

Anstelle der transversalen Komponenten $M_x(t)$ und $M_y(t)$ der Magnetisierung, ist es ebenso möglich die rein reelle asymmetrische Frequenzverteilung $p(\omega)$ zu betrachten. In vorangegangenen Arbeiten wurde diese Frequenzverteilung im Static-Dephasing-Regime von verschiedenen Autoren untersucht \cite{Case87,Seppenwoolde05,Durney89,Cutillo96,Ailion92,Christman96}. Hier weist die Frequenzverteilung eine asymmetrische Form mit scharfen Peaks bei $\omega=-\eta\delta\omega$ und $\omega=+2\eta\delta\omega$ auf, wie in Abb. \ref{Fig:Fig2} dargestellt. Außerdem ist die lokale Resonanzfrequenz für verschwindende Diffusion aufgrund des Zählers des Dipolfelds in Gl. \eqref{eq:frequenz} auf den Bereich $-\delta\omega\leq\omega\leq+2\delta\omega$ beschränkt. Der Ausdruck der Magnetisierung im Static-Dephasing-Grenzfall stimmt mit vorausgegangenen Ergebnissen überein, wie oben gezeigt \cite{Cheng01,Ziener07a}. Die minimale Resonanzfrequenz liegt am Äquator des Dipolfelds vor (bei $\theta=\pi/2$), während die maximale Resonanzfrequenz am Nord- oder Südpol des Dipolfelds für $\theta=0$ bzw. $\theta=\pi$ auftritt. Sobald der Diffusionseffekt berücksichtigt werden muss, erlaubt die Bewegung der Spins, dass Resonanzfrequenzen außerhalb des Intervalls $-\delta\omega\leq \omega \leq +2\delta\omega$ auftreten. Dieser Umstand erschwert die Berechnung des freien Induktionszerfalls mittels einer Fourier-Transformation der Frequenzverteilung gemäß Gl. \eqref{eq:RueckFT}, da die Intervallgrenzen $-\infty<\omega<+\infty$ Probleme bei der numerischen Evaluation verursachen können. Diese Probleme lösen sich durch die Methode, die in dieser Arbeit beschrieben wird, da ein expliziter Ausdruck der Magnetisierung in Gl. \eqref{eq:Result} angegeben ist. Für praktische Anwendungen ist es daher sinnvoll Gl. \eqref{eq:Fouriertransform} mit der Funktion $H$ aus Gl. \eqref{eq:H} für die Berechnung der Frequenzverteilung zu benutzen, während der Signalverlauf mit Gl. \eqref{eq:Result} berechnet werden sollte.

Darüber hinaus konnte gezeigt  werden, dass die Frequenzverteilung im Diffusions-Regime näherungsweise eine verschobene Lorentz-Kurve ist, wie in Abb. \ref{fig:ApproxFS} dargestellt. Die Frequenzverschiebung $\omega_\text{max}$ sowie die Breite der Frequenzverteilung hängen von der Position der Nullstelle in der komplexen Ebene ab. Ähnliche Ergebnisse für eine verschobene Lorentz-förmige Frequenzverteilung wurden in vorausgegangenen Arbeiten im Static-Dephasing-Regime erhalten \cite{Yablonskiy94,Cheng01}. In der vorliegenden Arbeit ist das Ergebnis jedoch im Diffusions-Regime für kleine Werte von $\taudom$ gültig. 

Die geometrische Beschreibung des Gewebes erlaubt es, zusammen mit der Strong-Collision-Näherung das Frequenzspektrum des Lungengewebes zu berechnen (siehe Abb. \ref{fig:Mulkern}). Insbesondere an den Flanken des Peaks stimmen das theoretische und das experimentelle Frequenzspektrum gut überein. Die Abweichung am Maximum des Peaks ergeben sich daraus, dass entsprechend dem Fourier-Theorem kleine Frequenzen zu großen Zeiten korrespondieren und für große Zeiten das Signal-Rausch-Verhältnis abnimmt.

\chapter{Danksagung}
Bei der Anfertigung dieser Bachelorarbeit wurde ich von vielen Personen unterstützen, wofür ich sehr dankbar bin. Zunächst möchte ich mich bei Christian Ziener bedanken. Er hat mir sehr dabei geholfen bei den Rechnungen das Wesentlich im Auge zu behalten und hat mich in vielen gemeinsamen Stunden stets motiviert und auch über diese Arbeit hinausgehend unterstützt. Ebenso möchte ich mich bei Felix Kurz und Thomas Kampf bedanken. Von ihnen habe ich wertvolle Anregungen zum Inhalt und der Gestaltungsweise dieser Bachelorarbeit bekommen.
Zudem möchte ich mich bei meinen Eltern Maria und Wolfgang Buschle und bei meiner Freundin Tina Kaufmann bedanken, die diese Arbeit zur Korrektur gelesen haben.
Zuletzt bedanke ich mich ausdrücklich bei Prof. Bachert und Prof. Schad für die Bereitschaft die vorliegende Arbeit zu begutachten.

\appendix
\chapter{Anhang}
\section{Static-Dephasing-Grenzfall}
\label{AnhangA}
\begin{figure}
\begin{center}
\includegraphics[width=9 cm]{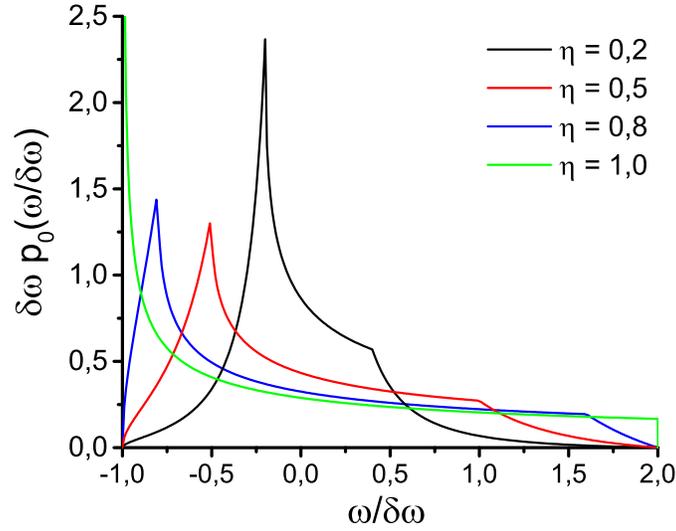}
\caption{
Frequenzverteilung $p_0(\omega)$ im Static-Dephasing-Grenzfall ($\taudom \to \infty$) für verschiedene Werte des Volumenverhältnis $\eta$. Auf der Abszisse bzw. Ordinate sind die dimensionslosen Größen $\omega /\delta\omega$ bzw. $\delta\omega p_0(\omega/\delta\omega)$ aufgetragen. Die lokale Larmor-Frequenz nimmt aufgrund des Zählers im Dipolfeld in Gl. \eqref{eq:frequenz} nur Werte zwischen $-\delta\omega \leq \omega \leq +2\delta\omega$ an (siehe Gl. \eqref{eq:p0}). Im Grenzfall $\eta \to 1$ kann die Frequenzverteilung mittels Gl. \eqref{eq:p0etaGrenz} berechnet werden (grüne Linie). Die Position und die Höhe der beiden auffälligen Peaks sind in Gl. \eqref{peak1} und Gl. \eqref{peak2} angegeben.
\label{fig:p0eta}
}
\end{center}
\end{figure}

Die kugelspezifische $h$-Funktion, welche die Zeitentwicklung der Magnetisierung im Static-Dephasing-Grenzfall in Gl.  \eqref{M01} bestimmt, wurde in \cite{Ziener07a} gefunden:  
\begin{equation} 
\label{app:kleinh}
h(x) = -\text{e}^{-\text{i} x} \, \frac{1+\I}{2}\sqrt{\frac{\pi}{6x}}\, \mathrm{erf}\left([\I-1]\sqrt{\frac{3x}{2}}\right) + x \! \int\limits_0^1  \!\! \text{d}z \, \left[3z^2 - 1\right] \left[ \text{Si}\left(x\left[3z^2 - 1\right] \right) - \text{i} \, \text{Ci} \left(x |3z^2 - 1|\right) \right] \,.
\end{equation}
Dabei sind Integralsinus und Integralcosinus folgendermaßen definiert:
\begin{align} 
\text{Si}(x) & = +\int_0^x \frac{\sin (t)}{t} \text{d}t \\
\text{Ci}(x) & = - \int_x^{\infty} \frac{\cos (t)}{t} \text{d}t \,.
\end{align}
Der Verlauf von $\text{Ci}(z)+\I \text{Si}(z)$ in der komplexen $z$-Ebene nennt sich Nielsen-Spirale.
Zudem bestimmt die Frequenzverteilung $p_0(\omega)$ im Static-Dephasing-Grenzfall die Magnetisierung $M_0(t)$ in Gl. \eqref{FT} und wurde in vorausgegangen Arbeiten bestimmt \cite{Cheng01,Ziener07a}:
\begin{align} \label{eq:p0}
p_0(\omega) & = \left\{
\begin{array}{ll}
\frac{\eta}{3 \sqrt{3} [1-\eta]} \:  \frac{\delta \omega}{\omega^2} \left[ 2-\frac{\omega}{\delta \omega}\right] \sqrt{1+\frac{\omega}{\delta \omega}} & \text{für }  \omega \in [-\delta \omega, -\eta \delta \omega ] \; \cup \; [+2\eta \delta \omega,+2 \delta \omega]\\[2ex]
\frac{\eta}{3 \sqrt{3} [1-\eta]} \:  \frac{\delta \omega}{\omega^2} \left[ \left[ 2-\frac{\omega}{\delta \omega}\right] \sqrt{1+\frac{\omega}{\delta \omega}} - \left[2-\frac{\omega}{\eta \, \delta \omega}\right] \sqrt{1+\frac{\omega}{\eta \, \delta \omega}} \right] & \text{für } \omega \in [ - \eta \delta \omega , +2\eta \delta \omega] \\[2ex]
0 &  \text{sonst} \,.
\end{array}
\right.
\end{align}
Die Frequenzverteilung ist in Abb. \ref{fig:p0eta} für verschiedene Werte des Volumenverhältnis $\eta$ dargestellt. Im Grenzfall $\eta \to 1$ ergibt sich die Frequenzverteilung zu
\begin{equation}
\label{eq:p0etaGrenz}
\lim_{\eta \to 1} p_0(\omega) = \frac{1}{2 \sqrt{3} \delta\omega \sqrt{\frac{\omega}{\delta \omega}+1}} \text{ für } -\delta\omega \leq \omega \leq +2\delta\omega
\end{equation}
und ist in der grünen Linie in Abb. \ref{fig:p0eta} dargestellt.
Durch die Ausführung der Fourier-Transformation (siehe Gl. \eqref{FT}) von Gl. \eqref{eq:p0etaGrenz} erhält man einen expliziten Ausdruck für die Magnetisierung:
\begin{align}
\lim_{\eta \to 1} M_0(t) &= -\E^{-\I \delta\omega t} \sqrt{\frac{\pi}{6\delta\omega t}} \frac{1+\I}{2} \text{erf}\left([\I-1] \sqrt{\frac{3}{2}\delta\omega t} \right)\\
&=+\E^{-\I \delta \omega t} \sqrt{\frac{\pi}{6\delta\omega t}} \left[\text{C}\left(\sqrt{\frac{6\delta\omega t}{\pi}}\right)+ \I \text{S}\left(\sqrt{\frac{6\delta\omega t}{\pi}}\right) \right]\,,
\end{align}
mit der Gaußschen Fehlerfunktion $\text{erf}(x)$. Die Funktionen $\text{C}(u)$ und $\text{S}(u)$ bezeichnen die Fresnelschen Integrale:
\begin{align}
&\text{C}(u) = \int \limits_{0}^{u} \cos \left(\frac{\pi}{2} x^2 \right) \D x\\
&\text{S}(u) = \int \limits_{0}^{u} \sin \left(\frac{\pi}{2} x^2 \right) \D x\,.
\end{align}
Der Verlauf von $\text{C}(u) + \I \text{S}(u)$ in der komplexen Ebene wird als Cornu-Spirale oder Klothoide bezeichnet. Die Magnetisierung weist somit in diesem Grenzfall einen spiralförmigen Verlauf in der komplexen Ebene auf. Dies kann wie in Abb. \ref{fig:Spiral} gezeigt, auf den Allgemeinfall übertragen werden.

In Abb. \ref{Fig:Fig2} stimmt die Frequenzverteilung für $\tau\delta\omega\to\infty$ (grüne Linie) mit der Frequenzverteilung im Static-Dephasing-Grenzfall gegeben in Gl. \eqref{eq:p0} überein. Die beiden Peaks liegen in diesem Grenzfall bei
\begin{align}
\label{peak1}
p_0(\omega=-\eta\delta\omega) & = \frac{1}{\delta\omega} \frac{2+\eta}{3\sqrt{3} \eta \sqrt{1-\eta}}\\
\label{peak2}
p_0(\omega=+2\eta\delta\omega) & = \frac{1}{\delta\omega} \frac{1}{6\eta} \sqrt{\frac{1+2\eta}{3}} \,.
\end{align}
Im Grenzfall $\eta \to 1$ wandert der linke Peak gegen unendlich, während der rechte Peak gegen $1/[6\delta\omega]$ konvergiert:
\begin{align}
&\lim_{\eta \to 1} p_0(\omega=-\delta\omega) = +\infty\\
&\lim_{\eta \to 1} p_0(\omega=+2\delta\omega) = \frac{1}{6\delta\omega} \,.
\end{align}

Zudem kann im Static-Dephasing-Grenzfall das Frequenzspektrum $\sigma_0(\omega)$, das intrinsische $T_2$-Relaxation berücksichtigt, analytisch nach Gl. \eqref{eq:Faltung} berechnet werden:
\begin{align}
\label{eq:FaltungSD}
\scriptstyle{\sigma_0(\omega)=\frac{\eta  T_2}{9 \pi[1 -\eta] \left[1+T_2^2 \omega ^2\right]^2} \Big[3 \left[1+T_2^2 \omega ^2\right]\left[\frac{1}{\eta }-1 \right]
}&\scriptstyle{+\sqrt{\frac{12}{T_2 \delta \omega}}\text{Re}\left([T_2 \omega -\I]^2 [1 -\I T_2 \omega+2 \I T_2 \delta \omega] \sqrt{\I+T_2 [\delta \omega +\omega]} \text{arctanh}\left(\sqrt{\frac{3 T_2\delta \omega}{\I+T_2 [\delta \omega +\omega]}}\right)\right)}\\
\nonumber
&\scriptstyle{
 -\sqrt{\frac{12}{\eta ^3 T_2 \delta\omega}}\text{Re} \left([T_2 \omega-\I]^2 [1-\I T_2\omega +2\I T_2 \eta \delta\omega] \sqrt{\I+T_2[\eta\delta\omega+\omega]} 
  \text{arctanh}\left(\sqrt{\frac{3 T_2 \eta \delta\omega}{\I+T_2[\eta\delta\omega + \omega]}}\right)\right)}\Big]\,.
\end{align}
Diese Gleichung kann sehr nützlich sein, um theoretische und experimentelle Frequenzverteilungen zu vergleichen.
\section{Strong-Collision-Näherung}
\label{AnhangB}
Die Bloch-Torrey-Gleichung \eqref{eq:BT} beschreibt die Zeitentwicklung der lokalen Magnetisierung $m(\mathbf{ r}, t)$ in einem inhomogenen Magnetfeld unter dem Einfluss von Diffusionseffekten auf Spin-tragende Teilchen. Für ein zweidimensionales magnetisches Dipolfeld kann diese Gleichung mit einem Separationsansatz gelöst werden \cite{Ziener12a}, da die lokale Larmor-Frequenz $\omega(\mathbf{r}) \propto r^{-2}$ dieselbe Abhängigkeit vom Radius wie der zweidimensionale Laplace-Operator $\Delta \propto r^{-2}$ aufweist. Im dreidimensionalen Fall (siehe Gl. \eqref{eq:frequenz}) hingegen ist dies aufgrund einer unterschiedlichen Abhängigkeit der lokalen Larmor-Frequenz vom Radius nicht möglich: $\omega (\mathbf{r}) \propto r^{-3}$. Darum ist es zweckmäßig eine Näherung der Bloch-Torrey-Gleichung in Form der Strong-Collision-Näherung zu benutzen. Die Strong-Collision-Näherung erlaubt korrekte Aussagen in allen Diffusionsregimen, wie in \cite{Dickson11} demonstriert wurde. Die folgende Ausführung orientiert sich an bereits dargestellten Ergebnissen \cite{Bauer99a, Bauer99c}.

Um die Näherung durchzuführen, muss die Bloch-Torrey-Gleichung diskretisiert werden:
\begin{equation}
|m(t)\rangle = \E^{\left[\mathbf{R}+\I \mathbf{\Omega} \right] t} |m(0)\rangle \, ,
\end{equation}
mit $\mathbf{R}=D \Delta$ und $\mathbf{\Omega} = \omega(\mathbf{r}) \mathbf{1}$.
Der ursprüngliche Magnetisierungsvektor ist proportional zum Gleichgewichtseigenvektor und kann normalisiert werden zu $|m(0)\rangle = |0\rangle$. Damit kann das Signal aus Gl. \eqref{eq:superposition} folgendermaßen umgeschrieben werden:
\begin{equation}
\label{app:MBraKet}
M(t) = \langle{0|\E^{\left[\mathbf{R}+\I \mathbf{\Omega} \right] t}|0} \rangle \,.
\end{equation}
Da der Operator $\mathbf{R}$ nicht-hermitesch ist, wird er durch den einfacheren Operator eines Markov-Prozesses ersetzt, der aus der statistischen Physik bekannt ist. Ein Markov-Prozess hängt nur vom aktuellen Zustand ab und ist von vorangegangenen Zuständen unabhängig. Somit ersetzt der Operator $\mathbf{R}=\lambda [\mathbf{\Pi}-\mathbf{1}]$ mit $\mathbf{\Pi} = |0\rangle \langle 0|$ den Diffusionsprozess. Für die weitere Berechnung muss die Konstante $\lambda$ mit bekannten Parametern verknüpft werden. 
Dazu kann die Frequenz-Korrelationsverteilung $K(t) = \langle \omega(0) \omega(t) \rangle$, wie in \cite{Ziener08} definiert, betrachtet werden. Die Mean-Relaxation-Time-Näherung $K(t) = K(0)\, \E^{-t/\tau}$ erlaubt es, die Korrelationszeit $\tau$ folgendermaßen auszudrücken:
\begin{equation}
\label{app:deftau}
\tau = \int_0^{\infty} \frac{K(t)}{K(0)} \D t \,.
\end{equation}
Die Korrelationszeit $\tau$ wurde in \cite{Ziener06} berechnet und ist in Gl. \eqref{eq:tau} angegeben.
Die Definition der Korrelationszeit aus Gl. \eqref{app:deftau} wird genutzt, um die Frequenz $\omega(t)$ in Abhängigkeit der Operatoren $\mathbf{\Omega}$ und $\mathbf{\Pi}$ anzugeben. Dadurch ergibt sich
\begin{equation}
\tau = \int_0^{\infty} \frac{\langle 0| \mathbf{\Omega} \E^{\lambda \left[\mathbf{1}-\mathbf{\Pi}\right] t} \mathbf{\Omega} |0 \rangle}{\langle 0| \mathbf{\Omega}^2 |0 \rangle} \D t \,.
\end{equation}
Eine Berechnung dieses Integrals führt zu $\lambda = \tau^{-1}$. Durch Einsetzen dieses Ergebnisses in Gl. \eqref{app:MBraKet} erhält man
\begin{equation}
M(t) = \langle{0|\E^{\left[\tau^{-1}\left[\mathbf{\Pi}-\mathbf{1}\right] +\I \mathbf{\Omega} \right] t}|0} \rangle \,.
\end{equation}
Wie oben erwähnt, ist die Anwendung der Laplace-Transformation auf $M(t)$ (siehe Gl. \eqref{eq:Laplacetransform}) ein geeigneter Ansatz:
\begin{equation}
\label{app:M(s)}
\hat M(s) =  \left \langle{0\left|\frac{1}{s\mathbf{1}-\I\mathbf{\Omega}-\tau^{-1} [\mathbf{\Pi}-\mathbf{1}]}\right|0} \right\rangle \,.
\end{equation}
Mit der Operatoridentität $[\mathbf{A}+\mathbf{B}]^{-1} = \mathbf{A}^{-1}-\mathbf{A}^{-1}\mathbf{B}[\mathbf{A}+\mathbf{B}]^{-1}$ mit $\mathbf{A}=\left[s+\tau^{-1} \right]\mathbf{1} -\I \mathbf{\Omega}$ und $\mathbf{B}=-\tau^{-1} \mathbf{\Pi}$, erhält man aus Gl. \eqref{app:M(s)} den Ausdruck für $\hat M(s)$:
\begin{align}
\hat M(s) &= \left\langle0\left|\frac{1}{\left[s+\tau^{-1}\right]\mathbf{1}-\I\mathbf{\Omega}}\right|0 \right\rangle + \tau^{-1}\left\langle0\left|\frac{1}{\left[s+\tau^{-1}\right]\mathbf{1}-\I\mathbf{\Omega}}\right|0 \right\rangle \left\langle0\left|\frac{1}{s\mathbf{1}-\mathbf{R}-\I\mathbf{\Omega}}\right|0 \right\rangle\\[1ex]
&= \hat M_0(s+\tau^{-1})+\tau^{-1} \hat M_0(s+\tau^{-1}) \hat M(s)\,.
\end{align}
Damit kann Gl. \eqref{eq:LT} erhalten werden.

\section{Inverse Laplace-Transformation}
\label{AnhangC}

Die inverse Laplace-Transformation der Funktion $\hat G (s)$ aus Gl. \eqref{eq:Gdach} kann durch Anwendung von Mellins Inversionsformel berechnet werden:
\begin{equation}
\label{app:Mellin}
G(t) = \frac{1}{2\pi \I} \int\limits_{\text{Re}(s_0)+\varepsilon-\I\infty}^{\text{Re}(s_0)+\varepsilon+\I\infty} \hat{G} (s) \E^{st} \D s \,,
\end{equation}
wobei die Singularität $s_0$ durch die Nullstelle des Nenners $N(s)$ aus Gl. \eqref{eq:Nenner} gegeben ist. Wie gefordert wird die Integration entlang einer Achse parallel zur imaginären Achse durchgeführt, die rechts von der Singularität $s_0$ liegt. Da $\lim_{|y|\to \infty} H(y)=1$ verschwindet die Funktion $\hat{G}(s)$ auf der linken und rechten Halbebene für $|s|\to \infty$ (siehe Gl. \eqref{eq:Gdach}). Somit kann unter Berücksichtigung des Integralkerns $\E^{st}$ der inversen Laplace-Transformation der Integrationsweg in Gl. \eqref{app:Mellin} gemäß dem Jordanschen Lemma geschlossen werden. Dies ist schematisch in Abb. \ref{fig:Bromwich} dargestellt. Letztendlich kann Gl. \eqref{app:Mellin} in die folgende Form umgeschrieben werden:
\begin{equation}
\label{app:Jordanlemma}
\ointctrclockwise\limits_{ABCA} \hat{G}(s)\E^{st} \D s = \int\limits_A^B \hat{G}(s)\E^{st} \D s = 2\pi\I G(t)\,.
\end{equation}
Aufgrund der Eigenschaften der arccoth-Funktion weist die Funktion $H$ in Gl. \eqref{eq:H} eine Verzweigungslinie entlang der imaginären Achse von $-\I$ bis $+2\I$ auf. Folglich besitzt auch die Funktion $\hat{G}(s)$ eine Verzweigungslinie entlang der imaginären Achse von  $-\I\taudom$ bis $+2\I\taudom$. Das Ringintegral kann unter Benutzung des Residuensatzes ausgewertet werden:
\begin{align}
\label{app:residuen}
\ointctrclockwise\limits_{ABCA} \hat{G}(s)\E^{st} \D s \,\,\, + \ointclockwise\limits_{DEFGD} \hat{G}(s)\E^{st} \D s = 2\pi\I \Theta\left(\taudom_0 -\taudom\right) \text{res}\left(\hat{G}(s)\E ^{st} ; s_0\right)\,,
\end{align}
wobei die Singularität $s_0$ für $\taudom<\taudom_0$ auftritt, wie durch die Heaviside-Funktion $\theta(\taudom_0-\taudom)$ sichergestellt wird. Dies ist in Abb. \ref{fig:VerlaufNullstelle} dargestellt.

Die Kombination von Gl. \eqref{app:Jordanlemma} und Gl. \eqref{app:residuen} ergibt einen Ausdruck der Funktion $G(t)$:
\begin{equation}
\label{app:result}
G(t) = \Theta \left(\taudom_0 -\taudom\right)  \text{res}\left(\hat{G}(s)\E ^{st} ; s_0\right) - \frac{1}{2\pi\I} \ointclockwise\limits_{DEFGD} \hat{G}(s)\E ^{st} \D s \,.
\end{equation}
Das Residuum lässt sich folgendermaßen berechnen:
\begin{align}
\label{app:Residuum}
\nonumber
\mathrm{res} \left(  \hat{G}(s)\E ^{st}  ; s_0 \right) &= \E ^{s_0 t} s_0 \frac{1-\eta}{N^{\prime}(s_0)}\\
&= \frac{ \E ^{s_0 t} s_0  \eta [1-\eta] \taudom}{\eta [1-\eta] \taudom - H^{\prime} \left( \frac{s_0}{\eta \taudom} \right) + \eta^2 H^{\prime} \left( \frac{s_0}{\taudom} \right)} \,.
\end{align}
Die Integration um die Verzweigungslinie kann in vier Teile zerlegt werden:
\begin{align}
\label{app:IntegralAufteilung}
\ointclockwise\limits_{DEFGD} \hat{G}(s)\E ^{st} \D s = \int\limits_D^E \hat{G}(s)\E ^{st} \D s + \int\limits_E^F \hat{G}(s)\E ^{st} \D s + \int\limits_F^G \hat{G}(s)\E ^{st} \D s +\int\limits_G^D \hat{G}(s)\E ^{st} \D s\,.
\end{align}
Folgende Teilstücke des Integrationsweges liefern keinen Beitrag:
\begin{equation}
\label{app:IntegralEnden}
\int\limits_E^F \hat{G}(s)\E ^{st} \D s = 0 = \int\limits_G^D \hat{G}(s)\E ^{st} \D s \,.
\end{equation}
Die verbleibenden Integrale können kombiniert und mit der Substitution $s=\I\taudom y$ vereinfacht werden:
\begin{align}
\label{app:Verzweigung_Gdiff}
\int\limits_{D}^{E} \hat{G}(s)\E ^{st} \D s+\int\limits_{F}^{G} \hat{G}(s)\E ^{st} \D s = \I\taudom \lim_{\varepsilon \to 0} \int\limits_{-1}^{+2} \left[\hat{G}(\taudom [\I y-\varepsilon])-\hat{G}(\taudom [\I y+\varepsilon])\right] \E ^{\taudom \I yt}\D y \,.
\end{align}
In Übereinstimmung mit Gl. \eqref{eq:Gdach} und Gl. \eqref{eq:Nenner} können die Grenzwerte $\lim_{\varepsilon \to 0} \hat{G}(\taudom [\I y\pm \varepsilon])$ in Abhängigkeit der Grenzwerte $\lim_{\varepsilon \to 0} H(\I y\pm \varepsilon)$ geschrieben werden. Um diese Grenzwerte zu bestimmen, ist es notwendig drei Fälle zu unterscheiden:
\begin{align}
\nonumber
&\underline{y>2\text{:}}\\
\label{app:Hgross}
& H(\I y) =  \frac{1}{3}+\frac{2}{3} \left[ 1 - \frac{2}{y}\right] \sqrt{\frac{1+y}{3}} \mathrm{arccoth} \left(\!\!\sqrt{\frac{1+y}{3}} \right) \in \mathbb{R} \\
\nonumber
&\underline{-1 \leq y \leq 2  \text{:}}\\
\label{app:Hmittel}
& \lim_{\varepsilon \to 0}\text{Re}(H(\I y+\varepsilon)) =  \frac{1}{3}+\frac{2}{3} \left[ 1 - \frac{2}{y}\right] \sqrt{\frac{1+y}{3}} \mathrm{arctanh} \left(\!\!\sqrt{\frac{1+y}{3}} \right) \\
\label{Imateil}
& \lim_{\varepsilon \to 0}\text{Im}(H(\I y+\varepsilon)) =  \frac{\pi}{3} \left[ 1 - \frac{2}{y}\right] \sqrt{\frac{1+y}{3}}\\
\label{Symm1}
& \lim_{\varepsilon \to 0}\text{Re}(H(\I y-\varepsilon))= + \lim_{\varepsilon \to 0}\text{Re}(H(\I y+\varepsilon))\\
\label{Symm2}
& \lim_{\varepsilon \to 0}\text{Im}(H(\I y-\varepsilon))= - \lim_{\varepsilon \to 0}\text{Im}(H(\I y+\varepsilon))\\
\nonumber
&\underline{y<-1 \text{:}}\\
\label{app:Hklein}
& H(\I y) =  \frac{1}{3}+\frac{2}{3} \left[ 1 - \frac{2}{y}\right] \sqrt{\frac{-1-y}{3}} \left[\frac{\pi}{2}- \mathrm{arctan}\left(\!\!\sqrt{\frac{-1-y}{3}} \right)\right] \in \mathbb{R} \,.
\end{align}
Mit den Symmetrieeigenschaften der $H$-Funktion aus Gl. \eqref{Symm1} und Gl. \eqref{Symm2} kann das verbleibende Integral in Gl. \eqref{app:Verzweigung_Gdiff} zu folgendem Integral vereinfacht werden: 
\begin{align}
\label{app:Verzweigung_Integrand}
\ointclockwise\limits_{DEFGD} \hat{G}(s)\E ^{st} \D s =
\lim_{\varepsilon \to 0} \int\limits_{-1}^{2} \!\!\frac{-2\I y [\taudom]^2[1-\eta] \text{Im} \left(\eta H(\I y+\varepsilon)-H\left(\frac{\I y+\varepsilon}{\eta}\right)\right) \E^{\I\taudom y t}}
{ \left[\text{Re} \left(\eta H(\I y+\varepsilon)-H \left(\frac{\I y+ \varepsilon}{\eta} \right)\right) +\I y\taudom[1-\eta] \right]^2 + \left[\text{Im} \left(\eta H(\I y+\varepsilon)-H \left(\frac{\I y+ \varepsilon}{\eta} \right)\right) \right]^2} \D y \,.
\end{align}
Durch Einsetzen der Gleichungen \eqref{app:Hgross}, \eqref{app:Hmittel}, \eqref{Imateil} und \eqref{app:Hklein} in Gl. \eqref{app:Verzweigung_Integrand} erhält man mit Gl. \eqref{app:result} den Ausdruck
\begin{align}
\label{Ergebnis-G}
\scriptstyle{G(t)} &= \frac{\scriptstyle{ \Theta \left( \taudom_0-\taudom \right)  9 [1-\eta] s_0^3 \mathrm{e}^{s_0 t}}}{\scriptstyle{3 [1-\eta]s_0[3 s_0+1]+ \eta\sqrt{3-\frac{3\mathrm{i} s_0}{\tau\delta\omega}} \left[s_0+\mathrm{i} \tau\delta\omega-\frac{3 [\tau\delta\omega]^2}{s_0+\mathrm{i} \tau\delta\omega}\right] \mathrm{arccoth}\left(\!\!\sqrt{\frac{1}{3}-\frac{\mathrm{i} s_0}{3 \tau\delta\omega}}\right)-\sqrt{3-\frac{3\mathrm{i} s_0}{\eta  \tau\delta\omega}}\left[s_0+\mathrm{i}\eta\tau\delta\omega-\frac{3 [\eta\tau\delta\omega]^2}{s_0+\mathrm{i} \eta  \tau\delta\omega}\right]\mathrm{arccoth}\left(\!\!\sqrt{\frac{1}{3}-\frac{\mathrm{i} s_0}{3 \eta\tau\delta\omega}}\right)}}\\
\nonumber
&-\!\!\int\limits^{-\eta}_{-1}\frac{\scriptstyle{\E ^{\I y\taudom t}\frac{\left[\taudom\right]^2}{3}  \left[1-\eta\right] \eta \left[2-y\right] \sqrt{\frac{1+y}{3}} \D y}}
{\scriptstyle{\frac{\pi^2}{9}\left[\eta \left[1-\frac{2}{y}\right] \sqrt{\frac{1+y}{3}}\right]^2+\left[ \left[\I y \taudom-\frac{1}{3}\right] \left[1-\eta\right]+\frac{2\eta}{3}\left[ 1-\frac{2}{y}\right] \sqrt{\frac{1+y}{3}} \mathrm{arctanh} \left(\!\!\sqrt{\frac{1+y}{3}} \right)
-\frac{2}{3} \left[1-\frac{2\eta}{y}\right] \sqrt{\frac{-1-\frac{y}{\eta}}{3}} \left[\frac{\pi}{2} - \mathrm{arctan}  \left(\!\!\sqrt{\frac{-1-\frac{y}{\eta}}{3}} \right) \right]\right]^2}}\\
\nonumber
&-\!\!\int\limits^{2\eta}_{-\eta}\frac{\scriptstyle{\E ^{\I y\taudom t}\frac{\left[\taudom\right]^2}{3}  \left[1-\eta \right] \left[ \eta \left[ 2-y\right] \sqrt{\frac{1+y}{3}}-\left[2\eta-y \right] \sqrt{\frac{1+\frac{y}{\eta}}{3}}\right]\D y}}
{\scriptstyle{\frac{\pi^2}{9} \left[\eta \left[1-\frac{2}{y}\right] \sqrt{\frac{1+y}{3}}-\left[1-\frac{2\eta}{y} \right] \sqrt{\frac{1+\frac{y}{\eta}}{3}}\right]^2+\left[\left[\I\taudom y -\frac{1}{3}\right] \left[1-\eta\right]+\frac{2\eta}{3}\left[1-\frac{2}{y}\right]\sqrt{\frac{1+y}{3}}\mathrm{arctanh}\left(\!\!\sqrt{\frac{1+y}{3}}\right)
-\frac{2}{3}\left[ 1-\frac{2\eta}{y} \right] \sqrt{\frac{1+\frac{y}{\eta}}{3}}\mathrm{arctanh}\left(\!\!\sqrt{\frac{1+\frac{y}{\eta}}{3}}\right)\right]^2}}\\
\nonumber
&-\!\!\int\limits^{2}_{2\eta}\frac{\scriptstyle{\E ^{\I y\taudom t}\frac{\left[\taudom\right]^2}{3}  \left[1-\eta\right] \eta \left[2-y\right] \sqrt{\frac{1+y}{3}}\D y}}
{\scriptstyle{\frac{\pi^2}{9}\left[\eta \left[1-\frac{2}{y}\right] \sqrt{\frac{1+y}{3}}\right]^2+\left[ \left[\I y \taudom-\frac{1}{3}\right] \left[1-\eta\right]+\frac{2\eta}{3}\left[ 1-\frac{2}{y}\right] \sqrt{\frac{1+y}{3}} \mathrm{arctanh} \left(\!\!\sqrt{\frac{1+y}{3}} \right)
-\frac{2}{3}\left[ 1-\frac{2\eta}{y} \right] \sqrt{\frac{1+\frac{y}{\eta}}{3}}\mathrm{arccoth}\left(\!\!\sqrt{\frac{1+\frac{y}{\eta}}{3}}\right)\right]^2}} \,.
\end{align}
Diese Gleichung ergibt mittels Gl. \eqref{eq:mf} die Zeitentwicklung der Magnetisierung, wie sie im Kapitel "`Ergebnisse"' in Gl. \eqref{eq:Result}  angegeben ist.

\newpage
\chapter*{Eigenständigkeitserklärung}
\addcontentsline{toc}{chapter}{Eigenständigkeitserklärung}

Ich versichere, dass ich diese Arbeit selbstständig verfasst und keine anderen als die angegebenen Quellen und Hilfsmittel benutzt habe.\\
\\
\\
\\

Heidelberg, den 16. März 2015

\end{document}